\documentclass[onefignum,onetabnum]{siamart220329}

\usepackage{lipsum}
\usepackage{amsfonts}
\usepackage{graphicx}
\usepackage{epstopdf}
\usepackage{algorithmic}

\ifpdf
  \DeclareGraphicsExtensions{.eps,.pdf,.png,.jpg}
\else
  \DeclareGraphicsExtensions{.eps}
\fi

\newsiamremark{remark}{Remark}
\newsiamremark{hypothesis}{Hypothesis}
\crefname{hypothesis}{Hypothesis}{Hypotheses}
\newsiamthm{claim}{Claim}
\newsiamthm{assum}{Assumption}
\newsiamthm{scheme}{Scheme}
\newsiamthm{lem}{Lemma}
\newsiamthm{thm}{Theorem}
\headers{Exact Conditional Score-Guided Generative Modeling for UQ}{Z.~Zhang, C.~Tatsuoka, D.~Xiu, G.~Zhang}

\title{
Exact Conditional Score-Guided Generative Modeling for Amortized Inference in Uncertainty Quantification\thanks{This manuscript has been authored by UT-Battelle, LLC, under contract DE-AC05-00OR22725 with the US Department of Energy (DOE). The US government retains and the publisher, by accepting the article for publication, acknowledges that the US government retains a nonexclusive, paid-up, irrevocable, worldwide license to reproduce the published form of this manuscript, or allow others to do so, for US government purposes. DOE will provide public access to these results of federally sponsored research in accordance with the DOE Public Access Plan.}
}

\author{
\hspace{0.6cm}Zezhong Zhang\thanks{Computer Science and Mathematics Division, Oak Ridge National Laboratory, Oak Ridge, TN (\email{zhangz2@ornl.gov})}
 \and 
Caroline Tatsuoka\thanks{Department of Mathematics, Ohio State University,
Columbus, OH (\email{xiu.16@osu.edu}, \email{tatsuoka.3@buckeyemail.osu.edu})}
\and
Dongbin Xiu\footnotemark[3]
\and\newline
Guannan Zhang\thanks{Computer Science and Mathematics Division, Oak Ridge National Laboratory, Oak Ridge, TN (\email{zhangg@ornl.gov}).}
}

\usepackage{amsopn}

\ifpdf

\hypersetup{
  pdftitle={},
  pdfauthor={Z.~Zhang, C.~Tatsuoka, D.~Xiu, G.~Zhang}
}
\fi

\usepackage{bbm}
\usepackage{amsmath, amssymb}
\usepackage{graphicx}
\usepackage{float}
\usepackage{booktabs}
\usepackage{array}
\usepackage{caption}
\usepackage{enumitem}
\usepackage{subcaption}

\newcommand{\mN}{\mathcal{N}}

\newcommand{\bE}{{\mathbb{E}}}

\newcommand{\bbR}{\mathbb{R}}

\newcommand{\bzero}{\mathbf{0}}

\newcommand{\bU}{\boldsymbol{U}}
\newcommand{\bV}{\boldsymbol{V}}
\newcommand{\bu}{\boldsymbol{u}}
\newcommand{\bv}{\boldsymbol{v}}

\newcommand{\bZ}{\boldsymbol{Z}}
\newcommand{\bz}{\boldsymbol{z}}
\newcommand{\bX}{\boldsymbol{X}}
\newcommand{\bx}{\boldsymbol{x}}
\newcommand{\bY}{\boldsymbol{Y}}
\newcommand{\sbY}{\text{\tiny \textbf{Y}}}
\newcommand{\by}{\boldsymbol{y}}

\newcommand{\bmm}{\boldsymbol{m}}
\newcommand{\bI}{\boldsymbol{\rm I}}
\newcommand{\bH}{\boldsymbol{\rm H}}
\newcommand{\bmu}{\boldsymbol{\mu}}
\newcommand{\bSigma}{\boldsymbol{\Sigma}}
\newcommand{\bsigma}{\boldsymbol{\sigma}}
\newcommand{\bveps}{\boldsymbol{\varepsilon}}

\newcommand{\bone}{\boldsymbol{1}}
\newcommand{\bW}{\boldsymbol{W}}
\newcommand{\bb}{\boldsymbol{b}}

\newcommand{\deq}{\stackrel{\text{d}}{=}}

\makeatletter
\DeclareRobustCommand{\cev}[1]{%
  \mathpalette\do@cev{#1}%
}
\newcommand{\do@cev}[2]{%
  \fix@cev{#1}{+}%
  \reflectbox{$\m@th#1\vec{\reflectbox{$\fix@cev{#1}{-}\m@th#1#2\fix@cev{#1}{+}$}}$}%
  \fix@cev{#1}{-}%
}
\newcommand{\fix@cev}[2]{%
  \ifx#1\displaystyle
    \mkern#23mu
  \else
    \ifx#1\textstyle
      \mkern#23mu
    \else
      \ifx#1\scriptstyle
        \mkern#22mu
      \else
        \mkern#22mu
      \fi
    \fi
  \fi
}

\usepackage{cite}
\usepackage{titlesec}
\usepackage{placeins} 
\usepackage{float}
\usepackage{caption}
\allowdisplaybreaks
\begin{document}

\maketitle

\begin{abstract}
%
We propose an efficient framework for amortized conditional inference by leveraging exact conditional score-guided diffusion models to train a non-reversible neural network as a conditional generative model. Traditional normalizing flow methods require reversible architectures, which can limit their expressiveness and efficiency. Although diffusion models offer greater flexibility, they often suffer from high computational costs during inference. To combine the strengths of both approaches, we introduce a two-stage method. First, we construct a training-free conditional diffusion model by analytically deriving an exact score function under a Gaussian mixture prior formed from samples of the underlying joint distribution. This exact conditional score model allows us to efficiently generate noise-labeled data, consisting of initial diffusion Gaussian noise and posterior samples conditioned on various observation values, by solving a reverse-time ordinary differential equation. Second, we use this noise-labeled data to train a feedforward neural network that maps noise and observations directly to posterior samples, eliminating the need for reversibility or iterative sampling at inference time. The resulting model provides fast, accurate, and scalable conditional sampling for high-dimensional and multi-modal posterior distributions, making it well-suited for uncertainty quantification tasks, e.g., parameter estimation of complex physical systems. We demonstrate the effectiveness of our approach through a series of numerical experiments.

\end{abstract}



\section{Introduction}\label{sec:intro} 
Amortized Bayesian inference methods for uncertainty quantification have advanced significantly in recent years, with a focus on recovering unknown posterior distributions from observed data in computationally efficient and amortized manner. Leveraging deep learning for its generalization capabilities, inference networks side-step iterative and potentially costly traditional procedures such as Markov Chain Monte Carlo sampling (MCMC) \cite{hastings1970monte} or standard variational inference methods \cite{dempster1977maximum, hoffman2013stochastic}; MCMC requires rerunning the algorithm for obtaining a posterior distribution for every data observation of interest while standard variational inference similarly requires repetitive runs of an expectation-maximization optimization procedure. 

Conditional amortized inference methods are of particular interest: given samples from a joint distribution, they aim to learn the posterior distribution conditioned on arbitrary observations across the observational data space. Simulation-based inference methods, such as neural likelihood estimation, neural ratio estimation, and neural posterior estimation (NPE), construct neural network-based surrogate models to efficiently amortize the estimation of the likelihood, likelihood ratio, or posterior density, see \cite{cranmer2020frontier} for a review. A particular example of NPE method is Conditional Variational Autoencoder \cite{kingma2013auto}, which employs an encoder network to approximate posterior distributions over latent variables conditioned on observations, and a decoder network to model the conditional likelihood, with training driven by maximization of the evidence lower bound for learning of global variational parameters across the dataset. Another approach for conditional density recovery is conditional optimal transport, which approximates distributions via learned transport maps; recent methods include nonparametric entropic Brenier maps, flow-matching-based amortized vector fields, and neural network approximations of conditional transport maps \cite{baptista2024conditional, generale2024conditional, wang2023efficient}.

Score-based diffusion models \cite{song2020score} are powerful generative models that have demonstrated strong performance on complex conditional sample generation tasks in domains such as conditional image synthesis \cite{wang2024integrating, sun2024provable, ho2022classifier, ramesh2022hierarchical, saharia2022photorealistic} and natural language processing \cite{li2022diffusion, yu2022latent, gong2022diffuseq,chen2022analog}. 
These models define a diffusion process that gradually corrupts samples from the target distribution through the continuous injection of small Gaussian noise. In the continuous-time limit, this corruption process is described by a forward stochastic differential equation (SDE). A neural network is then trained to learn the score function, which characterizes the denoising direction and defines the drift term in the corresponding reverse-time SDE. Sample generation is performed by simulating this reverse-time SDE with the learned score network as the drift, transforming noise into data.

To sample from a conditional distribution (or a posterior distribution in Bayesian inference), existing approaches typically start with a learned prior score network and incorporate observation information into the reverse-time SDE to steer the sampling process toward the posterior. Depending on how the observation is incorporated, several methods have been proposed. A straightforward approach adds a scaled likelihood score or a measurement-matching term directly into the reverse process \cite{jalal2021robust, song2020score, chung2022diffusion, song2023pseudoinverse}, to approximate the unknown score function of the reverse-time SDE corresponding to the posterior. An alternative line of work \cite{feng2023score, mardani2023variational} employs variational inference techniques to approximate the posterior with a simpler distribution, avoiding the need to explicitly construct the posterior score. Other methods \cite{sun2024provable, wu2024principled, dou2024diffusion, cardoso2023monte} use MCMC techniques or particle-based sequential Monte Carlo to approximate the distribution of the posterior reverse-time SDE, offering asymptotic convergence guarantees in the limit of infinite computational resources. In \cite{venkatraman2024amortizing}, the posterior distribution is obtained by fine-tuning the prior diffusion model using the relative trajectory balance loss.

Recently, training-free score-based methods have been developed for UQ of complex systems \cite{liu2024diffusion, liu2024training, liu2024diffusion, yin_et_al_2024}; these approaches analytically approximate the score function using Monte Carlo estimates based on available prior samples. Training-free approximation methods are advantageous for efficiently approximating the score function at each pseudo-temporal location during the reverse-time process.  While it is initially required to store the training samples for score function approximation at inference time, this memory burden can be alleviated by building a labeled dataset for training of an amortized inference network that maps normal samples to target posterior samples directly, removing the need to retain or fine-tune the original model \cite{fan2024genai4uq,lu2024diffusion,liu2024diffusion}.

In this work, we aim to extend and improve upon training-free, score-based diffusion models for conditional sample generation. We present an exact conditional score-guided generative diffusion model, analytically derived under the assumption of a Gaussian mixture model (GMM) prior, where each Gaussian component is centered at an individual prior joint sample. Combined with appropriate Gaussian assumptions in the observation model, this formulation enables exact evaluation of the conditional score function without the need for neural network training.
Using the exact conditional score-guided diffusion model, we generate a noise-labeled dataset to train an amortized inference network, which eliminates the need to store potentially memory-intensive prior samples. Once trained, this network allows efficient, amortized sampling from conditional posterior distributions of interest. 
Within the exact conditional score-guided diffusion model, the only source of approximation error in recovering the conditional posterior density of the GMM for a given observation comes from the discretization of the reverse-time dynamics. The quality of the approximation thus depends on both the number and the coverage of the available prior samples. This approach is particularly advantageous in high-dimensional settings where the true underlying conditional distribution is complex but the prior dataset is sparse, as the Gaussian mixture prior provides a smoothed representation that can bridge gaps between samples and yield a numerically stable and tractable approximation to the true underlying posterior distribution.

The paper is outlined as follows: in Section 2 we define the problem setting, while in Section 3 we analytically derive the exact score function under the appropriate assumptions. In Section 4 we present numerical examples highlighting the applicability of the method and in Section 5 we provide closing remarks. 

\section{Problem setting -- amortized conditional sampling}\label{sec:prob}
Estimating conditional probability distribution is a central task in machine learning, statistics, and scientific computing, with applications ranging from probabilistic inference to Bayesian posterior estimation. 
We are interested in building a conditional generative sampler for the conditional distribution:
\begin{equation}\label{eq:cond_dist}
    p_{\bU | \bV}(\bu | \bv) = \frac{p_{\bU, \bV}(\bu, \bv)}{p_{\bV}(\bv)},
\end{equation}
where $\bU  \in \mathbb{R}^{d_u}$, $\bV \in \mathbb{R}^{d_v}$ are two continuous random vectors of interest, $p_{\bU , \bV}(\bu , \bv)$ is the joint probability density function (PDF), $p_{\bV}(\bv)$ is the marginal PDF of $\bV$, and $p_{\bU | \bV}(\bu | \bv)$ is the target conditional PDF. 
Traditional conditional density estimation methods typically estimate $p_{\bU | \bV}(\bu | \bv)$ for a fixed $\bv$, requiring resampling whenever $\bv$ changes, which limits their efficiency in large-scale applications.
These methods become computationally infeasible as the dimensionality of $\bU$ and $\bV$ increases, limiting their applicability in real-world problems. Additionally, classical approaches often struggle with complex, multi-modal distributions, where handcrafted density models may fail to capture intricate dependencies.

An amortized conditional generative model $F_{\theta}: \bbR^{d_v} \times \bbR^{d_u} \to \bbR^{d_u}$ is a transport mapping, parameterized by $\theta$, of the following form:
\begin{equation}\label{eq:fmap-sec2}
    \bU|\bV \deq F_{\theta} (\bV, \bZ)\, \text{ with }\, \bV\sim p_{\bV}(\bv), \, \bZ \sim \mathcal{N}(\mathbf{0}, \mathbf{I}_{d_{u}}),
\end{equation}
where $\bZ$ follows the standard normal distribution in $\bbR^{d_u}$ and $\theta$ is the tuning parameter of $F_\theta$. Once the model is trained, it can be used to generate samples from a conditional distribution $p_{\bU | \bV}(\bu | \bv)$ in Eq.~\eqref{eq:cond_dist} with any value of $\bV \sim p_{\bV}(\bv)$. In this work, we are given a dataset consisting of i.i.d. samples of the joint distribution, i.e.,
\begin{equation}\label{eq:dataset}
    \mathcal{D}_{\rm train} = \{ (\bu_k, \bv_k): k = 1, \ldots, K\},
\end{equation}
where $ (\bu_k, \bv_k) \sim p_{\bU, \bV}(\bu, \bv)$ for $k = 1, \ldots, K$.

We are interested in using the conditional score-based diffusion models to construct the desired conditional transport map $F_\theta$ as defined in Eq.~\eqref{eq:fmap-sec2}. 
Existing methods typically construct the conditional generator by first learning a prior score network from prior samples, and then performing posterior sampling by incorporating a scaled likelihood term or a measurement matching term into the reverse process. This enables sampling from an approximate conditional distribution $p_{\bU|\bV}(\bu|\bv)$. 
However, we argue that the current conditional diffusion models face several fundamental challenges. First, learning an accurate prior score network for high-dimensional, multi-modal distributions is computationally expensive. Minibatch-based training often fails to balance local feature fidelity with global distributional accuracy, making it difficult to capture the correct mode ratios.
Second, the support of the learned prior is limited to the training data. If the conditioning value shifts the posterior into a low-probability region of the prior, the score network may produce unreliable outputs, which could lead to instability of the sampling process.
Third, posterior sampling via likelihood-guided refinement is typically inexact and requires careful tuning of the guidance strength. Improper weighting can lead to either underestimated uncertainty (i.e., too little variation in posterior samples) or failure to satisfy the measurement constraint.


\subsection{Amortized Bayesian inference}\label{sec:abi}
The objective of this work is to reformulate the amortized conditional sampling problem in Eq.~\eqref{eq:cond_dist} as an amortized linear Bayesian inference problem and to develop an exact conditional score-guided diffusion model for this linear Bayesian problem, which eliminates the need to train a prior score network. This subsection introduces how to convert the problem of estimating the conditional distribution $p_{\bU|\bV}(\bU|\bV)$ in Eq.~\eqref{eq:cond_dist} to an amortized Bayesian inference problem. The key idea is to define two new random vectors, $\bX$ and $\bY$, such that
\begin{equation}\label{eq:xy}
    \bX := \left(\bU \atop \bV \right) \, \text{ and }\, \bY := \bH \bX + \bveps_{\by},
\end{equation}
where 
$\bX \in \mathbb{R}^{d_x}$, with $d_x = d_u + d_v$, is the concatenation of $\bU$ and $\bV$ from Eq.~\eqref{eq:cond_dist}, 
$\bY \in \mathbb{R}^{d_y}$ is the observable, $\bH \in \mathbb{R}^{d_y \times d_x}$ is a known linear transformation, and $\bveps_{\by} \sim \mN(0, \bSigma_{\bY})$, with $d_y = d_v$, denotes the Gaussian observation noise. Then, the posterior distribution $p_{\bX|\bY}(\bx|\by)$ can be written by Bayes' rule as
\begin{equation}\label{eq:bayes_post}
    p_{\bX|\bY}(\bx|\by) \propto p_{\bY | \bX}(\by |\bx) p_{\bX}(\bx),
\end{equation}
where the prior distribution $p_{\bX}(\bx)$ is the joint distribution in Eq.~\eqref{eq:cond_dist}, i.e.,
\begin{equation}
    p_{\bX}(\bx) = p_{\bU, \bV}(\bu, \bv)
\end{equation}
and the likelihood function is defined by
\begin{equation}\label{eq:likelihood-pdf1}
    p_{\bY | \bX}(\by | \bx) \propto |\bSigma_{\bY}|^{-\frac{1}{2}} \exp\left(-\frac{1}{2} (\bH \bx - \by)^\top \bSigma_{\bY}^{-1} (\bH \bx - \by) \right).
\end{equation}
To relate the posterior distribution $p_{\bX|\bY}(\bx|\by)$ to the target conditional distribution $p_{\bU | \bV}(\bu | \bv)$, we can choose 
\begin{equation}\label{eq:Hmat}
    \bH = [\bzero, \bI_{d_v}] \, \text{ such that }\, \bY = \bV + \bveps_{y},
\end{equation}
where $\bI_{d_v}$ denotes the identity matrix of size $d_v$ by $d_v$. This indicates that the $\bU$ component of the posterior distribution $p_{\bX|\bY}(\bx|\by)$ converges to the conditional distribution $p_{\bU | \bV}(\bu | \bv)$ as the observation noise $\bveps_{y}$ approaches zero, i.e.,
\begin{equation}\label{eq:limit}
     p_{\bX^{\bU}|\bY}(\bx^{\bU} | \by) \, \to \,  p_{\bU | \bV}(\bu | \bv) \; \text{ as }\; \bveps_{y} \rightarrow 0,
\end{equation}
where $p_{\bX^{\bU}|\bY}(\bx^{\bU} | \by)$ denotes $\bU$-component of $p_{\bX|\bY}(\bx | \by)$ defined in Eq.~\eqref{eq:xy}. 
Therefore, the Bayesian posterior distribution $p_{\bX^{\bU}|\bY}(\bx^{\bU} | \by)$ is a good approximation to the target conditional distribution $p_{\bU | \bV}(\bu | \bv)$ for a small $\bveps_{y}$ in Eq.~\eqref{eq:Hmat}. 
In Section \ref{sec:method}, we treat the Bayesian posterior $p_{\bX|\bY}(\bx | \by)$ in Eq.~\eqref{eq:bayes_post} as the target distribution and construct an exact conditional score-guided diffusion model to draw samples from the it for any given value $\by \sim p_{\bY}(\by)$, which will then be used to train the amortized conditional sampler.

\section{Conditional diffusion models with exact score computation}\label{sec:method}
This section introduces the main contribution of this paper, i.e., the derivation of the exact conditional score-guided diffusion model for sampling from the Bayesian posterior $p_{\bX|\bY}(\bx | \by)$ in Eq.~\eqref{eq:bayes_post}.
Section \ref{sec:prior} introduces a key assumption regarding the prior distribution $p_{\bX}(\bx)$. Section \ref{sec:diff} briefly overviews the score-based diffusion model used in this paper. Section \ref{sec:score} shows the derivation of the exact conditional score function. Section \ref{sec:special} presents a special case of the conditional score function when applied to sampling from the conditional distribution in Eq.~\eqref{eq:cond_dist}. 

\subsection{The prior distribution}\label{sec:prior}
Given the definition in Eq.~\eqref{eq:Hmat}, the samples in the training set in Eq.~\eqref{eq:dataset} becomes the samples of the prior distribution $p_{\bX}(\bx)$ in Eq.~\eqref{eq:bayes_post}, i.e.,
\begin{equation}\label{eq:dataset2}
    \mathcal{D}_{\rm train} = \{\bx_k: k = 1, \ldots, K\} \sim p_{\bX}(\bx).
\end{equation}
In real-world applications, the mathematical expression of the exact prior distribution $p_{\bX}(\bx)$ is usually unknown, so that we can only use the information provided by the dataset $\mathcal{D}_{\rm train}$. We observe that the samples in $\mathcal{D}_{\rm train}$ define an empirical distribution approximating the exact prior distribution. However, the finite support of the empirical distribution limits its ability to accurately represent the posterior, especially in high-dimensional settings.
To address this limitation, we impose the following assumption on the prior distribution.
\begin{assum}\label{thm:assump}
    The prior distribution $p_{\bX}(\bx)$ in Eq.~\eqref{eq:bayes_post} is defined by a Gaussian mixture model based on the dataset $\mathcal{D}_{\rm train}$ in Eq.~\eqref{eq:dataset2}, i.e., 
    \begin{equation}\label{eq:initial-dist}
    p_{\bX}(\bx) = \sum_{k=1}^K \pi_k p_{\bX|\xi}(\bx | k) \quad \text{with} \quad p_{\bX|\xi}(\bx | k) = \phi(\bx| \bmu_k,\bSigma_k),
\end{equation}
where $\phi(\bx| \bmu_k,\bSigma_k)$ is the Gaussian density function with mean $\bmu_k$ and covariance matrix $\bSigma_k$, and 
$\xi \in \{1,...,K\}$ is the indicator random variable determining the Gaussian component with the probability distribution $\mathbb{P}(\xi=k) = \pi_k = 1/K$.
\end{assum}

In this work, we set the number of Gaussian components $K$ in Eq.~\eqref{eq:initial-dist} equal to the number of samples in $\mathcal{D}_{\rm train}$ and assign uniform weights to all components, i.e., $\pi_k = 1/K$ for all $k$. 
In other words, the definition in Eq.~\eqref{eq:initial-dist} can be viewed as a Gaussian kernel density estimation of the unknown exact prior distribution. 
When substituting the prior from Eq.~\eqref{eq:initial-dist} into the linear Bayesian problem in Eq.~\eqref{eq:xy}, the posterior distribution $p_{\bX|\bY}(\bx | \by)$ in Eq.~\eqref{eq:bayes_post} is also a Gaussian mixture distribution. In theory, this means that a generative model is not strictly necessary, as sampling from the Gaussian mixture posterior is well understood. However, drawing samples from the Gaussian mixture posterior requires storing the entire training dataset $\mathcal{D}_{\rm train}$ and evaluating Gaussian kernel functions over the whole dataset, which is computationally expensive. Therefore, a compact, lightweight conditional generative model is highly desirable to reduce the complexity of generating new conditional samples given different conditioning value $\by$.

\subsection{Overview of the conditional score-based diffusion model}\label{sec:diff}
A score-based diffusion model consists of a forward process that transforms a target distribution into pure noise, and a reverse process that maps the noise back to the samples from the target distribution.
Let $\bZ_t^{\sbY} \in \bbR^{d_x}$ denote the forward process defined over the standard pseudo-temporal domain $t \in [0,1]$ with the initial distribution $\bZ_0^{\sbY}$ being the posterior distribution $p_{\bX|\bY}(\bx | \by)$ of interest from Eq.~\eqref{eq:bayes_post}. 
Following \cite{song2020score,kingma2021variational}, we can construct the forward process with the following linear SDE:
\begin{equation}\label{eq:forward}
d \bZ_t^{\sbY} = b(t) \bZ_{t}^{\sbY} dt + \sigma(t) d\bW_t\; \text{ with }\; \bZ_0^{\sbY} \deq \bX|\bY
\end{equation}
where $\bW_t \in \bbR^{d_x}$ is standard Brownian motion. The SDE coefficient is chosen to be 
\begin{equation}\label{eq:cof}
\begin{aligned}
b(t) = \frac{{\rm d} \log \alpha_t}{{\rm d} t} \;\;\; \text{ and }\;\;\; \sigma^2(t) = \frac{{\rm d} \beta_t^2}{{\rm d}t} - 2 \frac{{\rm d}\log \alpha_t}{{\rm d}t} \beta_t^2.
\end{aligned}
\end{equation}
which yields the following forward diffusion kernel:
\begin{equation}\label{eq:ab}
p_{Z_t|Z_0} = \phi(\alpha_t z_0 , \beta^2_t \quad \text{with} \quad \alpha_t = 1-t, \;\; \beta_t^2 = t \;\; \text{ for } \;\; t \in [0,1].
\end{equation}
The choice of $\alpha_t$ and $\beta_t^2$, as suggested in \cite{JCP_2024_Score, ensf_cmame}, ensures that the terminal distribution $\bZ_1^{\sbY}$ approximates a standard Gaussian for all values of $\bY$. Unlike the traditional diffusion models, which typically handle a fixed prior, in the conditional setting, the entire forward process $\bZ_t^{\sbY}$ will depend on the conditioning value $\bY$ through the initial distribution $\bZ_0^{\sbY} \sim p_{\bX|\bY}(\bx | \by)$.

We further denote the score function for the forward SDE as
\begin{equation}\label{eq:score-def}
 S_{\bZ_t^{\sbY}}(\bz_{t}) = \nabla_{\bz_t} \log p_{\bZ_t^{\sbY}}({\bz}_t),
\end{equation}
which is the key component in constructing the reverse process. 

It is shown in \cite{song2020score} that the reverse process, which shares the same marginal PDF as the forward process $\bZ_t^{\sbY}$, can be obtained through a probability flow ordinary differential equation (ODE) going from standard Gaussian distribution at $t=1$ to the target posterior distribution at $t=0$. The reverse ODE is given by:
\begin{equation}\label{DM:RSDE}
d{\bZ}_t^{\sbY} = \left[ b(t){\bZ}_t^{\sbY} - \frac{1}{2}\sigma^2(t) S_{\bZ_t^{\sbY}}(\bZ_t^{\sbY})\right] dt \; , \; \bZ_1^{\sbY} \deq \mN(0, \bI)
\end{equation}
where $S_{\bZ_t^{\sbY}}(\cdot)$ is the score function of the forward SDE defined in Eq.~\eqref{eq:score-def}.

As noted in \cite{liu2024diffusion}, for a fixed $\by$, this ODE defines a deterministic mapping between the standard Gaussian samples at $t=1$ and the posterior samples at $t=0$. Each posterior sample can be paired with its corresponding Gaussian noise for supervised training of an efficient sampler based on this mapping. In the conditional setting, the conditioning value $\by$ serves as an additional label for the mapping, enabling supervised learning of a conditional generator $F_{\theta}$ as described in Eq.~\eqref{eq:fmap-sec2}.

\subsection{Derivation of the exact score function}\label{sec:score}
Now, we derive the computable expression for the exact conditional score function $S_{\bZ_t^{\sbY}}(\cdot)$ in Eq.~\eqref{eq:score-def} for the Bayesian posterior distribution in Eq.~\eqref{eq:bayes_post} under Assumption \ref{thm:assump}. 
The derivation is achieved in two steps. The first step is to decompose the score function $S_{\bZ_t^{\sbY}}(\cdot)$ into several computable components, which will be given in Section \ref{sec:score_decomp}. The second step is to develop the computable expression of each of the components in the score decomposition, which will be given in Section \ref{sec:weight} and \ref{sec:condex}.

\subsubsection{Score decomposition}\label{sec:score_decomp}

We observe from Eq.~\eqref{eq:forward} that the dependence of $\bZ_t^{\sbY}$ on the observation $\bY$ is entirely through the initial distribution $\bZ_0^{\sbY} \sim p_{\bX|\bY}(\bx | \by)$. Since $p_{\bX|\bY}(\bx | \by)$ is unknown, we first link the posterior diffusion process $\bZ_t^{\sbY}$ with the prior diffusion process $\bZ_t$, where $\bZ_t$ denote the forward SDE with initial distribution $\bZ_0$ being the prior distribution $p_{\bX}(\bx)$. We can write 
\begin{equation}
    \begin{aligned}
       p_{\bZ_t^{\sbY}}(\bz_t|\by) 
      &= \int p_{\bZ^{\bY}_{t}|\bZ^{\bY}_{0}}(\bz_t | \bz_0, \by)  p_{\bZ^{\bY}_{0}} (\bz_0 | \by) d \bz_0 \\
      &= \int p_{\bZ_{t}|\bZ_{0}}(\bz_t | \bz_0)  p_{\bX|\bY}(\bz_0 | \by) d\bz_0\\
      & = \int p_{\bZ_{t}|\bZ_{0},\bY}(\bz_t | \bz_0, \by)  p_{\bZ_0|\bY} (\bz_0 | \by) d\bz_0\\
      &= p_{\bZ_t|\bY}(\bz_t | \by).
    \end{aligned}
\end{equation}
Here 
$p_{\bZ^{\sbY}_{t}|\bZ^{\sbY}_{0}} (\bz_t|\bz_0) = p_{\bZ_{t}|\bZ_{0}}(\bz_t|\bz_0) = p_{\bZ_{t}|\bZ_{0},\bY}(\bz_t|\bz_0, \by)$ because both $\bZ_t$ and $\bZ^{\sbY}_t$ shares the same transition kernel induced by the same stochastic dynamics, and the transition kernel is independent of the initial distribution, which is the only component affected by $\bY$.
Therefore, we now treat $p_{\bZ_t|\bY}(\cdot)$ as the target distribution, and its score function is defined by
\begin{equation}\label{eq:target-score-def}
    S_{\bZ^{\bY}_t}(\bz_t | \by) = S_{\bZ_t|\bY}(\bz_t | \by)  := \nabla_{\bz_t} \log p_{\bZ_t|\bY}(\bz_t) = \frac{\nabla_{\bz_t} p_{\bZ_t|\bY}(\bz_t|\by)}{p_{\bZ_t|\bY}(\bz_t |\by)}.
\end{equation}

Using the law of total probability to expand $p_{\bZ_t|\bY}(\bz_t | \by)$ through the latent variable $\xi$ defined in the Gaussian mixture initial distribution $\bZ_0 \sim p_{\bX}(\bx)$ in Eq.\eqref{eq:initial-dist}, we have 
\begin{equation}\label{eq:expand-by-xi}
    p_{\bZ_t|\bY}(\bz_t | \by) = \sum_{k=1}^K p_{\bZ_t|\bY,\xi}(\bz_t | \by, k)\, p_{\xi|\bY}(k | \by),
\end{equation}
where $p_{\xi|\bY}(k| \by) = \mathbb{P}(\xi=k|\bY=\by)$ is the probability of sampling from the $k$-th Gaussian component based on the conditional $\bY = \by$. 
Applying the Bayesian formula to $p_{\bZ_t|\bY, \xi}(\bz_t | \by, k)$ in Eq.\eqref{eq:expand-by-xi}, we then write $p_{\bZ_t|\bY}(\bz_t | \by)$ as 
\begin{align}
    p_{\bZ_t|\bY}(\bz_t | \by) 
    &= \sum_{k=1}^K  \frac{p_{\bY|\bZ_t, \xi}(\by| \bz_t, k)\, p_{\bZ_t | \xi} (\bz_t | k) }{ p_{\bY | \xi}(\by| k) }  p_{\xi|\bY}(k| \by) \\
    &= \sum_{k=1}^K  p_{\bY |\bZ_t, \xi}(\by | \bz_t, k) p_{\bZ_t | \xi} (\bz_t | k)  \frac{p_{\xi|\bY}(k | \by)}{ p_{\bY | \xi}(\by | k) }\\
    & = \sum_{k=1}^K  p_{\bY |\bZ_t, \xi}(\by | \bz_t, k) p_{\bZ_t | \xi} (\bz_t | k)  \frac{p_{\xi}(k)}{ p_{\bY }(\by) },
\end{align}
where the ratio ${p_{\xi | \bY}(k | \by)}/{ p_{\bY | \xi}(\by | k)}  = {p_{\xi}(k)}/{ p_{\bY }(\by) }$ is independent of $\bz_t$. Then, taking the gradient w.r.t. $\bz_t$ on both sides of the above equation, we have
\begin{equation}\label{eq:posterior-pdf-grad}
\begin{aligned}
    \nabla_{\bz_t}  p_{\bZ_t|\bY}(\bz_t | \by)
    = \sum_{k=1}^K &  \Bigg[ \nabla_{\bz_t} p_{\bY |\bZ_t, \xi}(\by | \bz_t, k) p_{\bZ_t | \xi} (\bz_t | k)\\
    & + p_{\bY |\bZ_t, \xi}(\by | \bz_t, k) \nabla_{\bz_t} p_{\bZ_t | \xi} (\bz_t | k)\Bigg] \frac{p_{\xi}(k)}{ p_{\bY }(\by) }.
\end{aligned}
\end{equation}
Substituting Eq.~\eqref{eq:posterior-pdf-grad} into the score function's definition in Eq.~\eqref{eq:target-score-def}, we have 
\begin{equation}\label{eq:final-score-1}
   \begin{aligned}
    S_{\bZ_t | \bY}(\bz_t | \by) 
     = &  \underbrace{\sum_{k=1}^K \frac{  p_{\bY |\bZ_t, \xi}(\by | \bz_t, k)   p_{\xi}(k)}{ p_{\bZ_t | \bY}(\bz_t | \by) p_{\bY}(\by)} \nabla_{\bz_t} p_{\bZ_t | \xi} (\bz_t | k)}_{I_1}\\
    & + \underbrace{\sum_{k=1}^K \frac{ p_{\bZ_t | \xi} (\bz_t | k) p_{\xi}(k)}{ p_{\bZ_t | \bY}(\bz_t | \by) p_{\bY}(\by) } \nabla_{\bz_t} p_{\bY |\bZ_t, \xi}(\by | \bz_t, k)}_{I_2},
\end{aligned} 
\end{equation}

Now we take a closer look at the two terms $I_1$ and $I_2$ in Eq.~\eqref{eq:final-score-1}. To proceed, we introduce the following Proposition on the score function of Gaussian distributions. 
\begin{proposition}\label{lemma:transition-kernel}
    If the initial state $\bZ_0$ of the diffusion model in Eq.~\eqref{eq:forward} follows Gaussian distribution 
    $\bZ_0 \sim \mN(\bmu, \bSigma)$ with the mean $\bmu$ and covariance matrix $\bSigma$, then the forward marginal distribution $p_{\bZ_t}(\bz_t)$ is a Gaussian distribution given by
    \begin{equation}\label{eq:forward-kernel}
    p_{\bZ_t}(\bz_t) = \phi(\bz_t; \bmu \alpha_t, \beta^2_t \bI_{d_x} + \alpha^2_t \bSigma),
    \end{equation}
    where $\phi(\cdot; \bmu, \bSigma)$ denotes the Gaussian PDF with mean $\bmu$ and covariance $\bSigma$. Moreover, the reverse transitional kernel $p_{\bZ_0 | \bZ_t}(\bz_0| \bz_t)$ is also a Gaussian distribution, i.e.,
    \begin{equation}\label{eq:reverse-kernel}
    p_{\bZ_0 | \bZ_t}(\bz_0| \bz_t) = \phi(\bz_0; \bmu_{0|t}(\bz_t,t), \bSigma_{0|t}(t)),
    \end{equation}
where $\bmu_{0|t}(\bz_t,t)$ and $\bSigma_{0|t}(t)$ is the conditional mean and covariance matrix of $\bZ_0 | \bZ_t$, respectively. Specifically, $\bmu_{0|t}(\bz_t,t)$ and $\bSigma_{0|t}(t)$ are given as
\vspace{-0.1cm}
\begin{align}
    \bmu_{0|t}(\bz_t)  &= \bmu + \alpha_t \bSigma (\alpha^2_t \bSigma + \beta^2_t \bI_{d_x})^{-1}(\bz_t - \bmu \alpha_t) \label{eq:reverse-mean} \\
    \bSigma_{0|t} &= \bSigma - \alpha^2_t \bSigma(\alpha^2_t \bSigma + \beta^2_t \bI_{d_x})^{-1} \bSigma \label{eq:reverse-cov} ,
\end{align}
where $\alpha_t$ and $\beta_t$ are given in Eq.~\eqref{eq:ab}.
\end{proposition}

\paragraph{\bf Decomposition of $I_1$ in Eq.~\eqref{eq:final-score-1} } 
Because $p_{\bZ_0 | \xi} (\cdot| k) = \phi(\cdot\,; \bmu_k, \bSigma_k)$ 
is a Gaussian distribution, Applying Proposition \ref{lemma:transition-kernel}
to the $k$-th Gaussian mode of the prior in Eq.~\eqref{eq:initial-dist}, we have that
$p_{\bZ_t | \xi} (\bz_t| k) = \phi(\bz_t; \bmu_k \alpha_t, \beta_t^2 {\rm I}_d + \alpha_t^2 \bSigma_k)$. From the gradient expression of the Gaussian PDF, we have 
\begin{equation}\label{eq:prior-cond-grad}
    \nabla_{\bz} p_{\bZ_t | \xi} (\bz_t|k) = p_{\bZ_t | \xi} (\bz_t | k) S_{\bZ_t|\xi}(\bz_t, t| k),
\end{equation}
where 
\begin{equation}\label{eq:prior-score-gmm}
     S_{\bZ_t|\xi}(\bz_t| k) = -(\beta^2_t {\rm I}_d + \alpha^2_t \bSigma_k)^{-1}(\bz_t-\bmu_k \alpha_t)
\end{equation}
is the score function of the Gaussian distribution $ p_{\bZ_t | \xi} (\cdot| k)$. Substituting Eq.~\eqref{eq:prior-cond-grad} into $I_1$ from  Eq.~\eqref{eq:final-score-1}, we can rewrite $I_1$ as 
\begin{equation}\label{eq:I1-final}
    \begin{aligned}
I_1&=\sum_{k=1}^K \frac{  p_{\bY |\bZ_t, \xi}(\by | \bz_t, k)   p_{\xi}(k)}{ p_{\bZ_t | \bY}(\bz_t | \by) p_{\bY}(\by)} \nabla_{z} p_{\bZ_t | \xi} (\bz_t | k)\\
&= \sum_{k=1}^K \frac{  p_{\bY |\bZ_t, \xi}(\by | \bz_t, k)   p_{\bZ_t | \xi} (\bz_t|k) p_{\xi}(k) }{ p_{\bZ_t | \bY}(\bz_t | \by) p_{\bY}(\by)} S_{\bZ_t|\xi}(\bz_t | k) \\
&= \sum_{k=1}^K \frac{ p_{\xi, \bZ_t, \bY}(k, \bz_t, \by) }{ p_{\bZ_t,\bY}(\bz_t, \by)} S_{\bZ_t|\xi}(\bz_t| k) \\
&=\sum_{k=1}^K p_{\xi|\bZ_t, \bY}(k | \bz_t,\by) S_{\bZ_t|\xi}(\bz_t| k),
\end{aligned}
\end{equation}
where $S_{\bZ_t|\xi}(\bz_t| k)$ can be computed using the formula in Eq.~\eqref{eq:prior-score-gmm} and the calculation of the weighting probability $p_{\xi|\bZ_t, \bY}(k | \bz_t,\by)$ will be introduced in 
Section \ref{sec:weight}.

\paragraph{\bf Decomposition of $I_2$ in Eq.~\eqref{eq:final-score-1}} 
We first deal with the gradient term $\nabla_{z} p_{\bY |\bZ_t, \xi}(\by | \bz_t,k)$. By introducing $\bZ_0$ as an intermediate variable, we can write the density function $p_{\bY | \bZ_t,\xi}(\by | \bz_t,k)$ as 
\begin{equation}\label{eq:y-zt-xi-decom}
\begin{aligned}
p_{\bY | \bZ_t,\xi}(\by | \bz_t, k)
&= \int p_{\bY | \bZ_0, \bZ_t, \xi}(\by | \bz_0, \bz_t, k) p_{\bZ_0 | \bZ_t, \xi}(\bz_0 | \bz_t, k) d \bz_0\\
&= \int p_{ \bY | \bZ_0}( \by | \bz_0) p_{ \bZ_0 | \bZ_t, \xi}(\bz_0 | \bz_t,k) d \bz_0,
\end{aligned}
\end{equation}
where $p_{\bY | \bZ_0}(\cdot) \equiv p_{\bY | \bX}(\cdot)$ is the likelihood function defined in Eq.\eqref{eq:likelihood-pdf1} and the density function $p_{\bZ_0 | \bZ_t, \xi}(\cdot| \bz_t, k)$ is the reverse transition kernel conditioned on the $k$-th Gaussian mode of the prior.

Because $p_{\bZ_0 | \xi} (\bz_0| k) \sim \mN(\bmu_k, \bSigma_k)$ is a Gaussian distribution, Eq.\eqref{eq:reverse-kernel} in Proposition \ref{lemma:transition-kernel} implies that $p_{\bZ_0 | \bZ_t, \xi}(\bz_0| \bz_t, k)$ in Eq.~\eqref{eq:y-zt-xi-decom} is also a Gaussian distribution, i.e.,
\begin{equation}\label{eq:reserse-kernel-gmm}
    p_{\bZ_0 | \bZ_t, \xi}(\bz_0| \bz_t, k) = \phi(\bz_0; \bmu_{0|t,k}(\bz_t), \bSigma_{0|t,k}),
\end{equation}
where the conditional mean $\bmu_{0|t,k}(\bz_t)$ and conditional covariance $\bSigma_{0|t,k}$ are given by
\begin{equation}\label{eq:mu_sig_0tk}
  \begin{aligned}
    \bmu_{0|t,k}(\bz_t) & = \bmu_k + \alpha_t \bSigma_k (\alpha^2_t \bSigma_k + \beta^2_t \bI_d)^{-1}(\bz_t - \bmu_k\alpha_t), \\
    \bSigma_{0|t,k} &= \bSigma_k - \alpha^2_t \bSigma_k(\alpha^2_t \bSigma_k + \beta^2_t \bI_d)^{-1} \bSigma_k.
\end{aligned}  
\end{equation}

Because the gradient $\nabla_{\bz_t} p_{\bY |\bZ_t, \xi}(\by | \bz_t,k)$ in $I_2$ of Eq.~\eqref{eq:final-score-1} is taken with respect to $\bz_t$, which is contained in the conditional mean $\bmu_{0|t,k}(\bz_t)$ of the reverse kernel $p_{\bZ_0 | \bZ_t, \xi}(\bz_0| \bz_t, k)$, we use a parametrization trick to move the dependency on $\bz_t$ outside of the Gaussian density $p_{\bZ_0 | \bZ_t, \xi}(\bz_0| \bz_t, k)$ as follows
\begin{equation}\label{eq:change-var}
    (\bZ_0 | \bZ_t, \xi=k) \deq \bmu_{0|t,k}(\bz_t) + L_{0|t,k} \bveps
\end{equation}
where $\bveps$ is the standard Gaussian vector and $L_{0|t,k}$ is the Cholesky component of $\bSigma_{0|t,k}$. Then we can re-write $\nabla_{\bz_t} p_{\bY | \bZ_t,\xi}(y | \bz_t, k)$ as 
\begin{equation}\label{eq:grad-y-zt-k}
    \begin{aligned}
    & \nabla_{\bz_t} p_{\bY | \bZ_t,\xi}(y | \bz_t, k)\\
    =& \nabla_{\bz_t} \int p_{ \bY | \bZ_0}( \by | \bz_0) \phi(\bz_0; \bmu_{0|t,k}(\bz_t), \bSigma_{0|t,k}) d \bz_0\\
    =& \nabla_{\bz_t} \int p_{\bY | \bZ_0}(\by | \bmu_{0|t,k}(\bz_t) + L_{0|t,k} \bveps) \phi(\bveps; \bzero, \bI_{d_x}) d \bveps\\
    =& \textbf{J}_{0|t,k} \int S_{\bY | \bZ_0}(\by | \bmu_{0|t,k}(\bz_t) + L_{0|t,k} \bveps) p_{\bY | \bZ_0}(\by | \bmu_{0|t,k}(\bz_t) + L_{0|t,k} \bveps) \phi(\varepsilon; \bzero, \bI_{d_x}) d \bveps\\
    =&  \textbf{J}_{0|t,k} \int S_{\bY | \bZ_0}(\by | \bz_0) p_{\bY | \bZ_0}(\by | \bz_0) p_{\bZ_0 | \bZ_t, \xi}(\bz_0 | \bz_t, k) d \bz_0, 
\end{aligned}
\end{equation}
where $\textbf{J}_{0|t,k}$ is the Jacobian matrix of $\bmu_{0|t,k}(\cdot)$ defined by
\begin{equation}\label{eq:mu-jacobian}
    \textbf{J}_{0|t,k} = \alpha_t \bSigma_k (\alpha^2_t \bSigma_k + \beta^2_t \bI_d)^{-1}.
\end{equation}
And in Eq.\eqref{eq:grad-y-zt-k}, we also used the following proposition for the gradient of the likelihood function $p_{\bY | \bZ_0}(\cdot) \equiv p_{\bY | \bX}(\cdot)$ defined in Eq.\eqref{eq:likelihood-pdf1}.
\begin{proposition}\label{prop-likelihood-score}
    For likelihood function  $p_{\bY | \bX}(\cdot)$ defined in Eq.\eqref{eq:likelihood-pdf1}, its gradient is given by 
    \begin{equation}\label{eq:likelihood-grad}
    \nabla_{\bx} p_{\bY | \bX} (\by | \bx) = p_{\bY | \bX} (\by | \bx) S_{\bY|\bX} (\by | \bx) 
\end{equation}
where the likelihood score function is 
\begin{equation}\label{eq:likelihood-score-vanila}
    S_{\bY|\bX} (\by | \bx) := \nabla_{\bx} \log p_{\bY | \bX} (\by | \bx) = - \bH^\top \bSigma_{\bY}^{-1} (\bH \bx - \by).
\end{equation}
\end{proposition}
Substituting Eq.\eqref{eq:grad-y-zt-k} back into $I_2$ from Eq.\eqref{eq:final-score-1}, we have 
\begin{equation}\label{eq:I2-final}
    \begin{aligned}
    I_2 
    &= \sum_{k=1}^K \frac{ p_{\bZ_t | \xi} (\bz_t |k) p_{\xi}(k)}{ p_{\bZ_t | \bY}(\bz_t | \by) p_{\bY}(\by) } \nabla_{\bz_t} p_{\bY |\bZ_t, \xi}(\by | \bz_t, k)\\
    &= \sum_{k=1}^K \frac{ p_{\bZ_t , \xi} (\bz_t, k) }{ p_{\bZ_t , \bY}(\bz_t, \by) } \textbf{J}_{0|t,k} \int S_{\bY | \bZ_0}(\by | \bz_0) p_{\bY | \bZ_0}(\by | \bz_0) p_{\bZ_0 | \bZ_t, \xi}(\bz_0 | \bz_t, k) d \bz_0\\
    &= \sum_{k=1}^K \textbf{J}_{0|t,k} \int \frac{ p_{\bZ_t , \xi} (\bz_t, k) p_{\bY | \bZ_0}(\by | \bz_0) p_{\bZ_0 | \bZ_t, \xi}(\bz_0 | \bz_t, k) }{ p_{\bZ_t , \bY}(\bz_t, \by)}     S_{\bY | \bZ_0}(\by | \bz_0) d \bz_0\\
    &= \sum_{k=1}^K \textbf{J}_{0|t,k} \int \frac{ p_{\bZ_0, \xi, \bZ_t, \bY}(\bz_0,k, \bz_t, \by) }{ p_{\bZ_t , \bY}(\bz_t, \by)} S_{\bY | \bZ_0}(\by | \bz_0) d \bz_0\\
    &= \sum_{k=1}^K \textbf{J}_{0|t,k} \int p_{\bZ_0, \xi | \bZ_t, \bY}(\bz_0, k,| \bz_t, 
    \by)     S_{\bY | \bZ_0}(\by | \bz_0) d \bz_0\\
    &= \sum_{k=1}^K \textbf{J}_{0|t,k} p_{\xi | \bZ_t, \bY}(k | \bz_t, \by) \int \frac{p_{\bZ_0, \xi | \bZ_t, \bY}(\bz_0, k,| \bz_t, \by)}{ p_{\xi | \bZ_t, \bY}(k|\bz_t, \by)}  S_{\bY | \bZ_0}(\by | \bz_0) d \bz_0\\
    &= \sum_{k=1}^K \textbf{J}_{0|t,k} p_{\xi | \bZ_t, \bY}(k | \bz_t, \by) \int p_{\bZ_0|\xi, \bZ_t, \bY}(\bz_0, |k, \bz_t, \by)  S_{\bY | \bZ_0}(\by | \bz_0) d \bz_0\\
    &= \sum_{k=1}^K p_{\xi | \bZ_t, \bY}(k | \bz_t, \by) \textbf{J}_{0|t,k} \bE \left[S_{\bY | \bZ_0}(\bZ_0) | \xi=k, \bZ_t=\bz_t, \bY=\by \right]
\end{aligned}
\end{equation}
Combining the decomposition of $I_1$ given in Eq.~\eqref{eq:I1-final} and the decomposition of $I_2$ in Eq.~\eqref{eq:I2-final}, we have final representation of the decomposition of the score function in Eq.~\eqref{eq:final-score-1}, i.e., 
\begin{equation}\label{eq:final-score-2}
\begin{aligned}
 S_{\bZ_t | \bY}(\bz_t | \by) =  \sum_{k=1}^K &  p_{\xi | \bZ_t, \bY}(k | \bz_t, \by) \Big\{S_{\bZ_t|\xi}(\bz_t|k) \\
      & +\textbf{J}_{0|t,k} \bE [S_{\bY | \bZ_0}(\bZ_0) | \xi=k, \bZ_t=\bz_t, \bY=\by ] \Big\},
\end{aligned}
\end{equation}
where $S_{\bZ_t|\xi}(\bz_t|k)$ is given in Eq.~\eqref{eq:prior-score-gmm} and $\textbf{J}_{0|t,k}$ is given in Eq.~\eqref{eq:mu-jacobian}. 
The computation of the weighting probability $p_{\xi | \bZ_t, \bY}(k | \bz_t, \by)$ is given in Section \ref{sec:weight} and the computation of the conditional expectation $\bE [S_{\bY | \bZ_0}(\bZ_0) | \xi=k, \bZ_t=\bz_t, \bY=\by ]$ is given in Section \ref{sec:condex}.

\subsubsection{Computation of the weighting probability}\label{sec:weight}
This subsection provides the formula for computing the weighting probability $p_{\xi| \bZ_t, \bY}(k | \bz_t, \by)$ of the score decomposition in Eq.~\eqref{eq:final-score-2}. Applying the Bayes' rule to the probability $p_{\xi| \bZ_t, \bY}(k | \bz_t, \by)$, we have
\begin{equation}\label{eq:cond-weight-decon}
p_{\xi | \bZ_t, \bY}(k | \bz_t, \by) \propto p_{\bY| \bZ_t,\xi}(\by | \bz_t,k) p_{\xi | \bZ_t}(k | \bz_t),
\end{equation}
where the proportionality holds because $\sum_k p_{\xi | \bZ_t, \bY}(k | \bz_t,  \by) = 1$. 
Applying the Bayes' rule to the probability $p_{\xi | \bZ_t}(k | \bz_t)$ in Eq.\eqref{eq:cond-weight-decon}, we can obtain
\begin{equation}\label{eq:prior-score-weight}
p_{\xi | \bZ_t}(k | \bz_t) \propto p_{\bZ_t | \xi}(\bz_t | k) p_{\xi}(k) =  \pi_k \phi(\bz_t; \bmu_k \alpha_t, \beta_t^2 \bI_{d_x} + \alpha_t^2 \bSigma_k ),
\end{equation}
where $p_{\bZ_t | \xi}(\bz_t | k)$ is the forward marginal distribution in Eq.~\eqref{eq:forward-kernel} in Proposition \ref{lemma:transition-kernel} and 
the proportionality is from $\sum_k p_{\xi| \bZ_t}(k | \bz_t)=1$ and $p_{\xi}(k) = \pi_k$. 
We use the expression of $p_{\bY| \bZ_t,\xi}(\by | \bz_t,k)$ given in Eq.~\eqref{eq:y-zt-xi-decom} to compute it in Eq.~\eqref{eq:cond-weight-decon}.

%

Since in Eq.~\eqref{eq:y-zt-xi-decom}, both $p_{ \bY | \bZ_0}( \by | \bz_0)$ and $p_{ \bZ_0 | \bZ_t, \xi}(\bz_0 | \bz_t,k)$ are Gaussian-type functions, their product is also Gaussian. By completing the square and marginalizing this Gaussian integral, we obtain the following expression for $p_{\bY | \bZ_t,\xi}(\by | \bz_t, k)$:

\begin{equation}\label{eq:y-zt-xi-result}
\begin{aligned}
     & p_{\bY | \bZ_t,\xi}(\by | \bz_t, k)\\
    \propto & |\bSigma_{0|t,k}|^{-1/2} |\hat\bSigma_k|^{1/2} \exp\left(-\frac{1}{2} \left[\bmu_{0|t,k}^\top(\bz_t) \bSigma_{0|t,k}^{-1} \bmu_{0|t,k}(\bz_t) -\bmm_k^\top \hat\bSigma_k \bmm_k\right] \right),
\end{aligned}
\end{equation}
where the terms without k are dropped due to proportionality, $\bSigma_{0|t,k}$ and $\bmu_{0|t,k}$ are defined in Eq.~\eqref{eq:mu_sig_0tk}, $\hat\bSigma_k$ and $\bmm_k$ are defined by 
\begin{align}
    \hat\bSigma_k &= \left(\bH^\top \bSigma_{\bY}^{-1} \bH + \bSigma_{0|t,k}^{-1}\right)^{-1},\label{eq:cond-weight-sigma}\\
    \bmm_k &= \bH^\top \bSigma_{\bY}^{-1} \by + \bSigma_{0|t,k}^{-1} \bmu_{0|t,k}(\bz_t).\label{eq:cond-weight-m}
\end{align}

In summary, substituting Eq.~\eqref{eq:prior-score-weight} and  Eq.~\eqref{eq:y-zt-xi-result} into Eq.~\eqref{eq:cond-weight-decon}, we have final computable representation of the weighting function $p_{\xi | \bZ_t, \bY}(k | \bz_t, \by)$ in the score decomposition in Eq.~\eqref{eq:final-score-2}, i.e.,
%
\begin{equation}\label{eq:cond-weight-result}
\begin{aligned}
    p_{\xi | \bZ_t, \bY}(k | \bz_t, \by) \propto & \,\pi_k \phi\left(\bz_t; \bmu_k \alpha_t, \beta_t^2 \bI_{d_x} + \alpha_t^2 \bSigma_k \right) |\bSigma_{0|t,k}|^{-1/2} |\hat\bSigma_k|^{1/2} \\
     & \times \exp\left(-\frac{1}{2} \left[\bmu_{0|t,k}^\top(\bz_t) \bSigma_{0|t,k}^{-1} \bmu_{0|t,k}(\bz_t) -\bmm_k^\top \hat\bSigma_k \bmm_k \right]\right),
\end{aligned}
\end{equation}
where $\pi_k$, $\bmu_k$ and $\bSigma_k$ are given by the prior distribution defined in Eq.~\eqref{eq:initial-dist}, $\alpha_t$ and $\beta_t$ are given in Eq.~\eqref{eq:ab}, $\bSigma_{0|t,k}$ and $\bmu_{0|t,k}$ are defined in Eq.~\eqref{eq:mu_sig_0tk}, $\hat\bSigma_k$ and $\bmm_k$ are defined in Eq.~\eqref{eq:cond-weight-sigma} and Eq.~\eqref{eq:cond-weight-m}, respectively. Therefore, the weighting probability is fully computable without any approximation. A special case is given in Section \ref{sec:special} for the case of having diagonal covariance matrix $\bSigma_k$ such that all the matrix inversions in Eq.~\eqref{eq:cond-weight-result} can be avoided.

\subsubsection{Computation of the conditional expectation}\label{sec:condex}
This subsection provides the formula for computing 
the conditional expectation $\bE [S_{\bY | \bZ_0}(\bZ_0) | \xi=k, \bZ_t=\bz_t, \bY=\by ]$ of the score decomposition in Eq.~\eqref{eq:final-score-2}. 

Since $p_{\bY|\bX}(\by|\cdot) \equiv p_{\bY|\bZ_0}(\by|\cdot)$, from Eq.~\eqref{eq:likelihood-score-vanila} in Proposition \ref{prop-likelihood-score} we have its score function as
\begin{equation}\label{eq:likelihood-score}
    S_{\bY | \bZ_0}(\by|\bz_0) = \nabla_{\bz} \log p_{\bY | \bZ_0}(\bz_0) = -\bH^\top \bSigma_{\bY}^{-1}(\bH \bz_0 - \by),
\end{equation}
which is a linear function of $\bz_0$. And this ensures that the expectation with respect $\bZ_0$ can be brought inside the score function, i.e., 
\begin{equation}
    \bE [S_{\bY | \bZ_0}(\bZ_0) | \xi=k, \bZ_t=\bz_t, \bY=\by ] = S_{\bY | \bZ_0}(\bE [\bZ_0| \xi=k, \bZ_t=\bz_t, \bY=\by ]).
\end{equation}
The computation of the expectation of the score function becomes the calculation of the expectation of $\bZ_0| \bZ_t, \xi, \bY$. For simplicity, we use the following notation to represent the conditional mean, i.e.,
\begin{equation}\label{eq:condmean}
    \bmu_{0|t,\xi,\bY}(\bz_t, k, \by) := \bE [\bZ_0| \xi=k, \bZ_t=\bz_t, \bY=\by ].
\end{equation}

According to the Bayes' rule, the posterior distribution $p_{\bZ_0| \bZ_t, \xi, \bY}(\bz_0|\bz_t,k,\by)$ can be written as
\begin{equation}\label{eq:post_z_0}
    p_{\bZ_0| \bZ_t, \xi, \bY}(\bz_0|\bz_t,k,\by) \propto 
    p_{\bY| \bZ_0, \bZ_t, \xi}(\by|\bz_0, \bz_t,k) p_{\bZ_0 | \bZ_t, \xi}(\bz_0| \bz_t, k),
\end{equation}
where the prior distribution $p_{\bZ_0 | \bZ_t, \xi}(\bz_0| \bz_t, k) = \phi(\bz_0; \bmu_{0|t,k}(\bz_t), \bSigma_{0|t,k}$ is a Gaussian distribution given in Eq.~\eqref{eq:reserse-kernel-gmm}, and the likelihood $p_{\bY| \bZ_0, \bZ_t, \xi}(\by|\bz_0, \bz_t,k)$, defined in Eq.~\eqref{eq:likelihood-pdf1}, is independent of $\bZ_t$ and $\xi$ and is a linear function of $\bZ_0$. According to the Kalman filter theory, the posterior distribution is also a Gaussian distribution, and the mean of $\bZ_0| \bZ_t, \xi, \bY$ can be calculated by the Kalman formula, i.e., 
\begin{equation}\label{eq:post-mean-kalman}
\begin{aligned}
    &\bmu_{0|t,\xi,\bY}(\bz_t, k, \by) \\
    =& \,\bmu_{0|t,k}(\bz_t) +  \bSigma_{0|t,k} \bH^\top (\bH \bSigma_{0|t,k} \bH^{\top} + \Sigma_{\bY})^{-1} (\by - \bH \bmu_{0|t,k}(\bz_t)),
\end{aligned}
\end{equation}
where $\bmu_{0|t,k}(\bz_t)$ and $\bSigma_{0|t,k}$ are defined in Eq.~\eqref{eq:mu_sig_0tk}, $\bH$ and $\bSigma_{\bY}$ is defined in Eq.~\eqref{eq:xy}.

%
%
Substituting Eq.~\eqref{eq:post-mean-kalman} into Eq.~\eqref{eq:likelihood-score}, we can compute the desired conditional expectation $\bE [S_{\bY | \bZ_0}(\bZ_0) | \xi=k, \bZ_t=\bz_t, \bY=\by ]$ in Eq.~\eqref{eq:final-score-2}. To do this, we start by computing $\bH \bmu_{0|t,k,\bY}(\bz_t, \by) - \by$, i.e.,
\begin{equation}\label{eq:Hmu}  
\small
\begin{aligned}
    & \bH \bmu_{0|t,k,\bY}(\bz_t, \by) - \by \\
    = &\bH \bmu_{0|t,k}(\bz_t) +  \bH \bSigma_{0|t,k} \bH^\top (\bH \bSigma_{0|t,k} \bH^{\top} + \Sigma_{\bY})^{-1} (\by - \bH \bmu_{0|t,k}(\bz_t)) - \by\\
    = &- (\by -\bH \bmu_{0|t,k}(\bz_t)) + \bH \bSigma_{0|t,k} \bH^\top (\bH \bSigma_{0|t,k} \bH^{\top} + \Sigma_{\bY})^{-1} (\by - \bH \bmu_{0|t,k}(\bz_t))\\
    = &(-\bI_{d_x} + \bH \bSigma_{0|t,k} \bH^\top (\bH \bSigma_{0|t,k} \bH^{\top} + \Sigma_{\bY})^{-1} )(\by - \bH \bmu_{0|t,k}(\bz_t))\\
    = &(-(\bH \bSigma_{0|t,k} \bH^{\top} + \Sigma_{\bY}) + \bH \bSigma_{0|t,k} \bH^\top ) (\bH \bSigma_{0|t,k} \bH^{\top} + \Sigma_{\bY})^{-1} (\by - \bH \bmu_{0|t,k}(\bz_t))\\
    = &- \Sigma_{\bY} (\bH \bSigma_{0|t,k} \bH^{\top} + \Sigma_{\bY})^{-1} (\by - \bH \bmu_{0|t,k}(\bz_t))\\
    = &\Sigma_{\bY} (\bH \bSigma_{0|t,k} \bH^{\top} + \Sigma_{\bY})^{-1} (\bH \bmu_{0|t,k}(\bz_t) - \by).
\end{aligned}   
\end{equation}
Substituting the above equation into the definition of the likelihood score function in Eq.~\eqref{eq:likelihood-score}, we have
\begin{equation}\label{eq:cond-expectation}
    \begin{aligned}
    & S_{\bY | \bZ_0}(\bmu_{0|t,k,\bY}(\bz_t, \by)) \\
    =&\, -\bH^\top \bSigma_{\bY}^{-1}(\bH \bmu_{0|t,k,\bY}(\bz_t, \by) - \by) \\
    =&\, -\bH^\top \bSigma_{\bY}^{-1} \bSigma_{\bY} (\bH \bSigma_{0|t,k} \bH^{\top} + \Sigma_{\bY})^{-1} (\bH \bmu_{0|t,k}(\bz_t) - \by) \\
    =&\, -\bH^\top (\bH \bSigma_{0|t,k} \bH^{\top} + \bSigma_{\bY})^{-1} (\bH \bmu_{0|t,k}(\bz_t) - \by), 
\end{aligned}
\end{equation}
where $\bmu_{0|t,k}(\bz_t)$ and $\bSigma_{0|t,k}$ are defined in Eq.~\eqref{eq:mu_sig_0tk}, $\bH$ and $\bSigma_{\bY}$ is defined in Eq.~\eqref{eq:xy}.

\subsubsection{Summary of the exact score computation}\label{sec:score_summary}
Combining all the calculations in Section \ref{sec:score_decomp} to Section \ref{sec:condex}, we have the final computational formula for the exact score function needed to solve the reverse ODE in Eq.~\eqref{DM:RSDE}
\begin{equation}\label{eq:final-score-final}
    \small S_{\bZ_t | \bY}(\bz_t | \by) = \sum_{k=1}^K p_{\xi | \bZ_t, \bY}(k | \bz_t, \by) \left( S_{\bZ_t|\xi}(\bz_t|k) +\textbf{J}_{0|t,k} S_{\bY | \bZ_0}(\bmu_{0|t,k,\bY}(\bz_t, \by)) \right)
\end{equation}
where $p_{\xi | \bZ_t, \bY}(k | \bz_t, \by)$ is given in Eq.\eqref{eq:cond-weight-result}, $S_{\bZ_t|\xi}(\bz_t|k)$ is given in Eq.\eqref{eq:prior-score-gmm}, $\textbf{J}_{0|t,k}$ is given in Eq.\eqref{eq:mu-jacobian} and $S_{\bY | \bZ_0}(\bmu_{0|t,k,\bY}(\bz_t, \by))$ is given in Eq.\eqref{eq:cond-expectation}.

The derived score function enables sampling from the posterior under the GMM prior assumption. Our ultimate objective is to approximate samples from the true underlying conditional distribution defined in Eq.~\eqref{eq:cond_dist}. In the next section, we describe a setting in which the exact score function is employed to generate approximate samples from the target conditional distribution, as well as how to construct the amortized inference network for sampling of posterior distributions for conditions across the observation space.

\subsection{Applying the exact score to the amortized conditional inference}\label{sec:special}
The exact score representation derived in Section \ref{sec:score} is for the general amortized Bayesian inference problem defined in Eq.~\eqref{eq:xy} and Eq.~\eqref{eq:bayes_post}, i.e., the definition of the observation matrix $\bH$ in Eq.~\eqref{eq:xy} could be any linear transformation. 

To apply the derived score function to the target amortized conditional inference problem in Eq.~\eqref{eq:cond_dist}, we provide the score representation with a special $\bH$ matrix defined in Eq.~\eqref{eq:Hmat}. Moreover, 
we also simplify the computational formula in Eq.~\eqref{eq:final-score-final} by assuming the covariance matrices $\bSigma_{\bY}$ in Eq.~\eqref{eq:xy} to be spherical and $\bSigma_k$ in Eq.~\eqref{eq:initial-dist} are block spherical, i.e., 
%
\begin{equation}\label{eq:sphe_sigma}
    \bSigma_{\bY} := \sigma_{\bY}^2 \mathbf{I}_{d_y}\;\; \text{ and }\;\; 
    \bSigma_k := 
    \begin{bmatrix}
        \sigma_{\bU}^2 \bI_{d_u} & \bzero\\
        \bzero & \sigma_{\bV}^2 \bI_{d_v}
    \end{bmatrix},
\end{equation}
where $\sigma_{\bY} \in \bbR$ and denotes the standard deviation of the observational noise and $\sigma_{\bU}, \sigma_{\bV} \in \bbR$ is the constant standard deviation of the first $d_u$ and last $d_v$ dimensions of each Gaussian mode of the Gaussian mixture model in Assumption \ref{thm:assump}. Note that all Gaussian components now share the same covariance matrix. By construction from Eq.~\eqref{eq:xy}, we have and $d_x = d_u + d_v$ and $d_y = d_v$. In addition, we assume the component mean $\{\bmu_k\}$ from $\bX$ are constructed from the paired data of $(\bU, \bV)$, i.e., $\bmu_k = (\bu_k , \bv_k), k=1,...,K$. The purpose of showing the score function with simplified covariance matrices is to illustrate the efficiency of the proposed score representation in a commonly used computational setting.

With the simplified covariance matrices, the mean and covariance in the reverse transitional kernel $p_{\bZ_0 | \bZ_t, \xi}(\bz_0| \bz_t, k) = \phi(\bz_0; \bmu_{0|t,k}(\bz_t), \bSigma_{0|t,k})$ in Eq.~\eqref{eq:reverse-kernel} can be simplified by blocking based on first $d_u$ and last $d_v$ dimensions. With that, we first rewrite $\bz_t$ as $\bz_t \equiv (\bz_t^u, \bz_t^v)$, where $\bz_t^u$ and $\bz_t^v$ is the first $d_u$ and last $d_v$ dimensions of $\bz_t$ respectively. Then we have 
\begin{equation}\label{eq:mu-uv-split}
\bmu_{0|t,k}(\bz_t)= 
\begin{bmatrix}
    \bmu_{0|t,k}^u(\bz_t^u)\\
    \bmu_{0|t,k}^v(\bz_t^v)
\end{bmatrix}
= 
\begin{bmatrix}
    s_{1,t}^u \bu_k + s_{2,t}^u \bz_t^u\\
    s_{1,t}^v \bv_k + s_{2,t}^v \bz_t^v
\end{bmatrix},\;\;
\bSigma_{0|t,k} = 
    \begin{bmatrix}
    s_{3,t}^u \bI_{d_u} & \bzero \\
    \bzero & s_{3,t}^v \bI_{d_v}
    \end{bmatrix}
\end{equation}
where the scalars $\{s_{i,t}^u, s_{i,t}^v\}_{i=1}^3$ are given by
\begin{equation}\label{eq:s-all}
   \begin{aligned}
    s_{1,t}^u = \frac{\beta_t^2}{\alpha_t^2 \sigma_{\bU}^2 + \beta_t^2}, \quad s_{1,t}^v = \frac{\beta_t^2}{\alpha_t^2 \sigma_{\bV}^2 + \beta_t^2}\\
    s_{2,t}^u = \frac{\alpha_t \sigma_{\bU}^2}{\alpha_t^2 \sigma_{\bU}^2 + \beta_t^2} , \quad s_{2,t}^v = \frac{\alpha_t \sigma_{\bV}^2}{\alpha_t^2 \sigma_{\bV}^2 + \beta_t^2}\\
    s_{3,t}^u = \frac{\sigma_{\bU}^2 \beta_t^2}{\alpha_t^2 \sigma_{\bU}^2 + \beta_t^2} , \quad s_{3,t}^v = \frac{\sigma_{\bV}^2 \beta_t^2}{\alpha_t^2 \sigma_{\bV}^2 + \beta_t^2}
\end{aligned} 
\end{equation}
On the other hand, the Jacobian $\textbf{J}_{0|t,k}$ in Eq.~\eqref{eq:mu-jacobian} is simpflied to
\begin{equation}\label{eq:cond-sample-jacobian}
    \textbf{J}_{0|t,k} = 
    \begin{bmatrix}
    s_{2,t}^u \bI_{d_u} & \bzero \\
    \bzero & s_{2,t}^v \bI_{d_v}
    \end{bmatrix},
\end{equation}
which is also independent of $k$.

Moreover, we can simplify the weighting probability $p_{\xi| \bZ_t, \bY}(k | \bz_t, \by)$ in Eq.~\eqref{eq:cond-weight-result}. The matrix $\hat\bSigma_k$ in Eq.~\eqref{eq:cond-weight-sigma} can be simplified to
\begin{align}
    \hat\bSigma_k 
    &= \left(\bH^\top \bSigma_{\bY}^{-1} \bH + \bSigma_{0|t,k}^{-1}\right)^{-1} = 
    \begin{bmatrix}
    s_{3,t}^u \bI_{d_u} & \bzero \\
    \bzero & \left(\dfrac{\sigma_{\bY}^2 s_{3,t}^v }{\sigma_{\bY}^2 + s_{3,t}^v} \right)\bI_{d_v}.
    \end{bmatrix}
\end{align}
Then the mean vector $\bmm_k$ in Eq.~\eqref{eq:cond-weight-m} can be simplified to
\begin{align}
    \bmm_k 
    &=  \bH^\top \bSigma_{\bY}^{-1} \by + \bSigma_{0|t,k}^{-1} \bmu_{0|t,k}(\bz_t)
    =  \left[\dfrac{\bmu_{0|t,k}^u(\bz_t^u)}{s_{3,t}^u} \atop \dfrac{\bmu_{0|t,k}^v(\bz_t^v)}{s_{3,t}^v} + \dfrac{\by}{\sigma_{\bY}^2} \right],
\end{align}

Therefore, the computation in Eq.~\eqref{eq:y-zt-xi-result} can be simplified to 
\begin{equation}\label{eq:msigm}
    \begin{aligned}
    \bmm_k^\top \hat\bSigma_k \bmm_k
    &= s_{3,t}^u \left\|\frac{\bmu_{0|t,k}^u(\bz_t^u)}{s_{3,t}^u}\right\|_2^2 + \left(\frac{\sigma_{\bY}^2 s_{3,t}^v }{\sigma_{\bY}^2 + s_{3,t}^v} \right) \left\|\frac{\by}{\sigma_{\bY}^2}  + \frac{\bmu_{0|t,k}^v(\bz_t^v)}{s_{3,t}^v}\right\|_2^2,
\end{aligned}
\end{equation}
\begin{equation}\label{eq:musigmu}
  \begin{aligned}
    \bmu_{0|t,k}^\top(\bz_t) \bSigma_{0|t,k}^{-1} \bmu_{0|t,k}(\bz_t) 
    = \frac{\left\|\bmu^u_{0|t,k}(\bz_t^u)\right\|_2^2 }{s_{3,t}^u} + \frac{\left\|\bmu^v_{0|t,k}(\bz_t^v)\right\|_2^2 }{s_{3,t}^v} 
\end{aligned}  
\end{equation}
which leads to 
\begin{equation}\label{eq:musigm}
   \begin{aligned}
    & \bmu_{0|t,k}^\top(\bz_t) \bSigma_{0|t,k}^{-1} \bmu_{0|t,k}(\bz_t)  - \bmm_k^\top \hat\bSigma_k \bmm_k \\
    = &\, \frac{\|\bmu^v_{0|t,k}(\bz_t^v)\|_2^2}{s_{3,t}^v} - \left(\frac{\sigma_{\bY}^2 s_{3,t}^v }{\sigma_{\bY}^2 + s_{3,t}^v} \right) \left\|\frac{\by}{\sigma_{\bY}^2} + \frac{\bmu_{0|t,k}^v(\bz_t)}{s_{3,t}^v}\right\|_2^2 \\
    =&\, \frac{\|\bmu^v_{0|t,k}(\bz_t^v)\|_2^2}{s_{3,t}^v} - \left(\frac{\sigma_{\bY}^2 s_{3,t}^v }{\sigma_{\bY}^2 + s_{3,t}^v} \right) \left(\frac{\|\by\|_2^2}{\sigma_{\bY}^4} + \frac{ 2\by^\top \bmu_{0|t,k}^v(\bz_t^v)}{\sigma_{\bY}^2 s_{3,t}^v} + \frac{\| \bmu_{0|t,k}^v(\bz_t^v)\|_2^2}{(s_{3,t}^v)^2}\right) \\
    =&\, c_1 + \frac{1}{\sigma_{\bY}^2 + s_{3,t}^v} \left(\left(\frac{\sigma_{\bY}^2 + s_{3,t}^v}{s_{3,t}^v} - \frac{\sigma_{\bY}^2}{s_{3,t}^v}\right)\|\bmu^v_{0|t,k}(\bz_t^v)\|_2^2 - 2\by^\top \bmu_{0|t,k}^v(\bz_t^v) \right)\\
    = &\,c_1 + \frac{\|\bmu^v_{0|t,k}(\bz_t^v)\|_2^2 - 2\by^\top \bmu_{0|t,k}^v(\bz_t^v)}{\sigma_{\bY}^2 + s_{3,t}^v} \\
    = &\, c_2+ \frac{\|\bmu^v_{0|t,k}(\bz_t^v) - \by\|_2^2}{\sigma_{\bY}^2 + s_{3,t}^v}, 
\end{aligned} 
\end{equation}
where $c_1$ and $c_2$ represent the terms that are independent of $k$ that will be dropped later in Eq.~\eqref{eq:cond-sample-cond-weight}.
In addition, the forward marginal distribution is also simplified as 
\begin{equation}
    p_{\bZ_t | \xi} (\bz_t | k) = \phi\left( \bz_t; 
    \begin{bmatrix}
        \bu_k \alpha_t \\
        \bv_k \alpha_t
    \end{bmatrix}, 
    \begin{bmatrix}
        (\beta_t^2 + \alpha_t^2 \sigma_{\bU}^2) \bI_{d_u} & \bzero\\
        \bzero & (\beta_t^2 + \alpha_t^2\sigma_{\bV}^2) \bI_{d_v}
    \end{bmatrix} \right)
\end{equation}
Using the above expressions, the weighting probability $p_{\xi | \bZ_t, \bY}(k | \bz_t, y)$ simplifies to
\begin{equation}\label{eq:cond-sample-cond-weight}
    p_{\xi | \bZ_t, \bY}(k | \bz_t, \by) \propto \exp\left( 
    -\frac{||\bz_t^u - \alpha_t \bu_k ||_2^2}{\beta_t^2 + \alpha_t^2 \sigma_{\bU}^2} - 
    \frac{||\bz_t^v - \alpha_t \bv_k ||_2^2}{\beta_t^2 + \alpha_t^2 \sigma_{\bV}^2} - 
    \frac{||\bmu^v_{0|t,k}(\bz_t^v) - \by||_2^2}{\sigma_{\bY}^2 + s_{3,t}^v} 
    \right),
\end{equation}
where the term $c_2$ in Eq.~\eqref{eq:musigm} are dropped due to the proportionality, i.e., the sum of all the weighting probabilities is one $\sum_k p_{\xi | \bZ_t, \bY}(k | \bz_t, \by) = 1$.

Additionally, we can simplify the conditional expectation representation given in Eq.~\eqref{eq:cond-expectation} to the following form:
\begin{equation}
    \begin{aligned}
    &\bE [S_{\bY | \bZ_0}(\bZ_0) | \xi=k, \bZ_t=\bz_t, \bY=\by ]\\[2pt]
    =&\,S_{\bY | \bZ_0}(\bmu_{0|t,k,\bY}(\bz_t, \by))\\[3pt]
    =&-\bH^\top (\bH \bSigma_{0|t,k} \bH^{\top} + \Sigma_{\bY})^{-1} (\bH \bmu_{0|t,k}(\bz_t) - \by)\\
    =& - \frac{1}{ \sigma_{\bY}^2 + s_{3,t}^v } \bH^\top (\bH \bmu_{0|t,k}(\bz_t) - \by)\\
    =& -
    \begin{bmatrix}
        \bzero_{d_u}\\
        \dfrac{\bmu_{0|t,k}^v(\bz_t^v) - \by}{\sigma_{\bY}^2 + s_{3,t}^v}
    \end{bmatrix},
\end{aligned}
\end{equation}
where $\bzero_{d_u} \in \bbR^{d_u}$ is a vector of all zeros. 

Combining the above simplified expressions, the final score representation in Eq.~\eqref{eq:final-score-final} can be simplified to
\begin{equation}\label{eq:final_score_spherical}
    \begin{aligned}
     S_{\bZ_t | \bY}(\bz_t | \by) 
    = & \sum_{k=1}^K p_{\xi | \bZ_t, \bY}(k | \bz_t, \by) \left( S_{\bZ_t|\xi}(\bz_t|k) +\textbf{J}_{0|t,k} S_{\bY | \bZ_0}(\bmu_{0|t,k,\bY}(\bz_t, \by)) \right) \\
    = & \sum_{k=1}^K p_{\xi | \bZ_t, \bY}(k | \bz_t, \by) \begin{bmatrix}
    - \dfrac{\bz_t^u - \alpha_t \bu_k}{\beta_t^2 + \alpha_t^2 \sigma_{\bU}^2}\\
    - \dfrac{\bz_t^v - \alpha_t \bv_k}{\beta_t^2 + \alpha_t^2 \sigma_{\bV}^2} - s_{2,t}^v \dfrac{\bmu_{0|t,k}^v(\bz_t) - \by}{\sigma_{\bY}^2 + s_{3,t}^v}
    \end{bmatrix},
\end{aligned}
\end{equation}
where $\{s_{i,t}^u, s_{i,t}^v\}_{i=1}^3$ are given in Eq.~\eqref{eq:s-all} and $p_{\xi | \bZ_t, \bY}(k | \bz_t, \by)$ is given in Eq.\eqref{eq:cond-sample-cond-weight}. Here, $\by$ is the condition for $\bV$ in the conditional distribution $p_{\bU | \bV}(\bu | \by)$. 

We now describe how the amortized inference network is constructed; we consider our joint sample set of size $K$ defined in Eq.~\eqref{eq:dataset2}, $\{\bx_{k}\}_{k=1}^K$. 
Utilizing this dataset to define the GMM prior, we can now run the exact score-guided diffusion model with the score function defined in Eq.~\eqref{eq:final_score_spherical} to obtain the following labeled dataset: \begin{equation} \label{eq:labeled_dataset}
\mathcal{S}_{\rm label} = \left\{\left({\bu}^{(j)}, \mathbf{z}^{(j)}, \by^{(j)}\right): j = 1,\dots,J\right\},
\end{equation}
where ${\by}^{(j)}$ is a sample from the marginal distribution $p_{\bY}({\by})$ of the observational data and serves as the conditioning variable. The latent variable $\mathbf{z}^{(j)} \sim \mathcal{N}(0, \mathbf{I}_{d_u + d_v})$ is the Gaussian noise input to the reverse-time ODE. $\bu^{(j)} \sim p_{\bX^{\bU} | \bY}(\bx^{\bU}|{\by}^{(j)})$ is the $\bU$-component of the Bayesian posterior $p_{\bX | \bY}(\bx|{\by}^{(j)})$, and it is obtained by running the reverse ODE with initial noise $\mathbf{z}^{(j)}$ and conditioning variable ${\by}^{(j)}$, and then extracting the 
$\bU$ component of the final sample.

This labeled dataset is then used to train a fully-connected, feed-forward neural network ${F}_\theta: \mathbb{R}^{d_v}\times \mathbb{R}^{d_u+d_v} \rightarrow \mathbb{R}^{d_u}$ using a standard L2 loss 
\begin{equation}\label{eq:model-loss}
    \min_{\theta} \sum_j \| F_{\theta}(\by^{(j)}, \mathbf{z}^{(j)}) - \bu^{(j)} \|_2^2.
\end{equation}
Given an input observation ${\by}$ and Gaussian samples, the trained neural network $F_{\theta}(\cdot)$ can generate samples from the Bayesian posterior $p_{\bX^{\bU}|\bY}(\bx^{\bU}|{\by})$, following the sampling procedure described in Eq.\eqref{eq:fmap-sec2}. According to Eq.~\eqref{eq:limit}, this posterior distribution can serve as an approximation to the target conditional distribution $p_{\bU|\bV}(\bu|{\bv})$, provided that the hyperparameters $\sigma^2_{\bU}, \sigma^2_{\bV}$ and $\sigma^2_{\bY}$ are appropriately chosen.

\subsection{Discussion on selection of hyperparameters $\sigma^2_{\bU}, \sigma^2_{\bV}$ and $\sigma^2_{\bY}$}
The first numerical example illustrates how varying $\sigma^2_{\bU}$, $\sigma^2_{\bV}$, and $\sigma^2_{\bY}$ influences the error of the approximate distribution produced by the diffusion model in a 1-dimensional bimodal setting. We note the optimal choice of $\sigma^2_{\bU}$, $\sigma^2_{\bV}$ and $\sigma^2_{\bY}$ will be problem dependent. To simplify the parameter selection process in the presented experiments, we consider variance selection of the joint samples $\bX = [\bU, \bV]^\top$ by fixing $\sigma^2_{\bV} = \sigma^2_{\bU}$. As an initial selection of $\sigma^2_{\bU}$ , we choose to utilize the average squared distance quantity between joint samples and its nearest neighbor, normalized by the total number of dimensions $d_u + d_v$.  In the examples below, we select $\sigma^2_{\bU}$ on the same order of magnitude of this quantity. We note that when $\sigma^2_{\bU} \ll 1$ and $\sigma^2_{\bV} \ll 1$, the posterior distribution approaches a weighted sum of Dirac delta functions centered at the training sample locations; this may lead to memorization \cite{baptista2025memorization} and an under-smoothed approximation of the true underlying conditional distribution, so care must be taken to avoid such parameter settings. Due to the potential over-smoothing effects of $\sigma^2_{\bY}$ as demonstrated below in Table~\ref{tab:full_comparison} and Figure~\ref{fig:rows_of_table_comparisons1-3}, we select $\sigma^2_{\bY}$ to be $10^{-4}$ or $10^{-5}$ in the presented examples. A rigorous study of how to select these parameters is ongoing work.  

\section{Numerical Experiments}\label{sec:num}

In this section, we present numerical experiments to demonstrate the ability of the exact conditional score-guided diffusion model to perform uncertainty quantification across a variety of problems. For systems without inherent randomness, we introduce artificial uncertainty by adding white noise to the observed data. In our first example in Section~\ref{sec:ex1}, we perform a series of ablation studies to analyze sources of error and evaluate the impact of parameter selection on approximation accuracy. Section~\ref{sec:20D_2Mode_GMM} presents a 20-dimensional conditional distribution example, and Section~\ref{sec:elliptic_PDE} presents a parameter inference problem for a 2-dimensional elliptic PDE.  

\subsection{1-dimensional bimodal conditional distribution.}\label{sec:ex1}

We begin by considering a 2-dimensional non-Gaussian joint distribution $p_{\bU,\bV}(\bu, \bv)$ to evaluate the accuracy of the proposed method in approximating the bimodal conditional distribution $p_{\bU|\bV}(\bu| \bv)$. The joint distribution is defined as
\begin{equation}\label{bimodal_example}
\bU \sim \text{Uniform}[-2,2] ,\quad  \bV = \bU^2 + \bveps, \quad \text{and} \quad \bveps \sim \mathcal{N}(0, 0.1).
\end{equation}
We generate datasets of joint samples $\{(\bu_{k},\bv_{k})\}_{k=1}^K$ and consider two prior samples size: $K =500$ and $K =5000$. Following the previous notations introduced in Eq.~\eqref{eq:xy}, we define the associated Bayesian problem for the conditional DM as  $\bX:= [\bU, \bV]^\top$ and $\bY := [1,0]^\top \bX + \bveps_{\bY}$, where $\varepsilon_{\bY} \sim \mN(0, \sigma_{\bY}^2)$. For all experiments, we consider $p_{\bU | \bV}(\bu | 1)$ as the target conditional distribution, and approximate it using the conditional DM applied to the Bayesian formulation $p_{\bX | \bY}(\bx | 1)$.

To quantify approximation error,  we compute the Kullback–Leibler (KL) divergence between the samples generated from DM and several reference distributions:
\begin{equation}\label{eq_kl}
D_{\rm KL}(p_{\rm ref} || p_{\rm DM}) = \int p_{\rm ref}(x)\log\left(\frac{p_{\rm ref}(x)}{p_{\rm DM}(x)}\right),dx,
\end{equation}
where $p_{\rm DM}$ denotes the probability density function estimated via kernel density estimation (KDE) from samples generated by the exact conditional diffusion model. We consider the following three reference distributions to calculate the KL divergence: 

\begin{itemize}[left=0.5cm]
    \item Bayesian GMM posterior ($p_{\rm BGMM}$):
    This is the $\bU$ components from the Bayesian GMM approximation $p_{\bX|\bY}(\bx|\by)$. 
    Since this is the exact distribution the diffusion model is designed to sample from, this error $e_{\rm BGMM} := D_{KL}(p_{\rm BGMM} || p_{\rm DM})$ quantifies how well the conditional DM samples from its target. As the conditional score function is exact, this error should reflect only the numerical discretization error from solving the reverse-time ODE.
    \item GMM conditional distribution ($p_{\rm GMM}$): This is the direct conditional distribution $p_{\bU|\bV}(\bu|\bv)$ computed from the empirical GMM prior constructed from the joint samples $\{(\bu_k, \bv_k)\}_{k=1}^K$. The corresponding KL divergence $e_{\rm GMM} := D_{KL}(p_{\rm GMM} || p_{\rm DM})$ measures the divergence between the DM sample and the conditional distribution defined by the GMM. This includes both the discretization error and the error due to the Bayesian relaxation in Eq.~\eqref{eq:xy}, which is primarily determined by $\sigma^2_{\bY}$.
    \item The exact Posterior ($p_{\rm exact}$): 
    This is the true posterior from Eq.~\eqref{bimodal_example}, which can be computed analytically 
    \begin{equation*} 
    p_{\rm exact}(\bu|\bv) := \frac{p_{\bV|\bU}(\by|\bv)p_{\bU}(\bu)}{p_{\bV}(\bv)} = \frac{e^{-\frac{(\bv-\bu^2)^2}{2\sigma^2}} \mathbf{1}_{[-2,2]}(\bu)}{\int_{-\infty}^\infty e^{-\frac{(\bv-\bu^2)^2}{2\sigma^2}}d\bu}.
    \label{exact_posterior}
    \end{equation*} 
    The corresponding KL divergence $e_{\rm exact} := D_{KL}(p_{\rm exact} || p_{\rm DM})$ captures the total approximation error, including discretization error, Bayesian relaxation error, which is controlled by $\sigma^2_{\bY}$, and GMM approximation error, which is controlled by $\sigma^2_{\bU}, \sigma^2_{\bV}$. 
\end{itemize}

\vspace{0.2cm} 
\begin{center}
\small
\begin{tabular}{>{\small}c|ccc|ccc}
\hline
 & $K$ & $\sigma^2_{\bU}$ & $\sigma^2_{\bY}$ & $e_{\rm exact}$ & $e_{\rm GMM}$ & $e_{\rm BGMM}$\\
\hline
C1  & 500  & 0.005 & 0.0001 & 1.28e-01 & 2.80e-03 & 2.79e-03 \\
C2  & 500  & 0.01  & 0.0001 & 7.96e-02 & 2.25e-03 & 2.25e-03 \\
C3  & 500  & 0.05  & 0.0001 & 3.87e-01 & 1.30e-03 & 1.30e-03 \\
\hline
C4  & 5000 & 0.001 & 0.0001 & 2.86e-02 & 4.30e-03 & 3.77e-03 \\
C5  & 5000 & 0.005 & 0.0001 & 2.77e-02 & 2.00e-03 & 1.98e-03 \\
C6  & 5000 & 0.01  & 0.0001 & 6.45e-02 & 1.80e-03 & 1.78e-03 \\
\hline
C7  & 5000 & 0.005 & 0.001  & 2.90e-02 & 2.25e-03 & 1.97e-03 \\
C8  & 5000 & 0.005 & 0.01   & 4.40e-02 & 7.47e-03 & 1.63e-03 \\
C9  & 5000 & 0.005 & 0.1    & 2.60e-01 & 1.64e-01 & 1.13e-03 \\
\hline
\end{tabular}
\captionof{table}{KL divergences between diffusion model samples and various reference distributions for different combinations of $K$, $\sigma^2_{\bU}$, and $\sigma^2_{\bY}$, with $\sigma^2_{\bV} = \sigma^2_{\bU}$ and reverse-time ODE discretization step $\Delta \tau=10^{-3}$.
Results show that discretization error $e_{\rm BGMM}$ remains small and stable (see $C_1 - C_9$); the Bayesian approximation error $e_{\rm GMM}$ increases with $\sigma^2_{\bY}$ (see $C_7 - C_9$); and the overall error $e_{\rm exact}$ can be minimized by appropriately tuning $\sigma^2_{\bU}$ and increasing $K$ (see $C_1 - C_6$).
}
\label{tab:full_comparison}
\end{center}

In Table \ref{tab:full_comparison}, we report the computed KL divergences for various combinations of $K$, $\sigma^2_{\bU}$, and $\sigma^2_{\bY}$, while fixing $\sigma^2_{\bV} = \sigma^2_{\bU}$ for simplicity. The pseudo-time domain is discretized using a step size of $\Delta \tau = 10^{-3}$. Several key observations can be drawn from the results: 
Firstly, $e_{\rm BGMM}$ remains constantly on the order of $10^{-3}$ across all configurations ($C_1–C_9$), confirming that the ODE discretization error is small and stable due to the fine step size. This supports the validity of our exact conditional score computation and also indicates that this discretization does not significantly contribute to other approximation errors.  
Secondly, for $e_{\rm GMM}$, we observe from cases $C_7-C_9$ that the error increases with larger values of $\sigma^2_{\bY}$. This behavior reflects the impact of the Bayesian reformulation in Eq.~\eqref {eq:xy}, where the $\bveps_{\bY} \sim \mN(0, \sigma^2_{\bY} \bI)$ introduces additional uncertainty into the inference process. Notably, this error can be made arbitrarily small by selecting sufficiently small values for $\sigma^2_{\bY}$, thereby recovering the behavior of the exact conditional distribution. 
Thirdly, comparing $e_{\rm exact}$ across configurations $C_1 - C_6$, where both $e_{\rm BGMM}$ and $e_{\rm GMM}$ (from small $\bsigma^2_{\bY}$) are already negligible,  we find that the dominant sources of error arise from the finite sample size $K$ and the GMM prior variance $\bsigma^2_{\bU}$. For a fixed $K$, the error is minimized at an intermediate value of $\bsigma^2_{\bU}$; both overly small and overly large values lead to suboptimal performance (as seen by comparing $C_2$ with $C_1$ and $C_3$, or $C_5$ with $C_4$ and $C_6$). Furthermore, increasing $K$ from 500 to 5,000 results in a smaller optimal $\bsigma^2_{\bU}$ with a better $e_{\rm exact}$ (comparing $C_2$ to $C_5$). 
This trend suggests that with more prior samples, the GMM posterior can better approximate the true conditional distribution, and the smoothing parameters $(\sigma^2_{\bU}, \sigma^2_{\bV})$ should be reduced accordingly. A smaller $(\sigma^2_{\bU}, \sigma^2_{\bV})$ enables the GMM posterior to rely more heavily on the available dataset rather than the smoothing imposed by the selected Gaussian covariance.
\begin{figure}[h!]
    \centering
\includegraphics[width=0.8\linewidth]{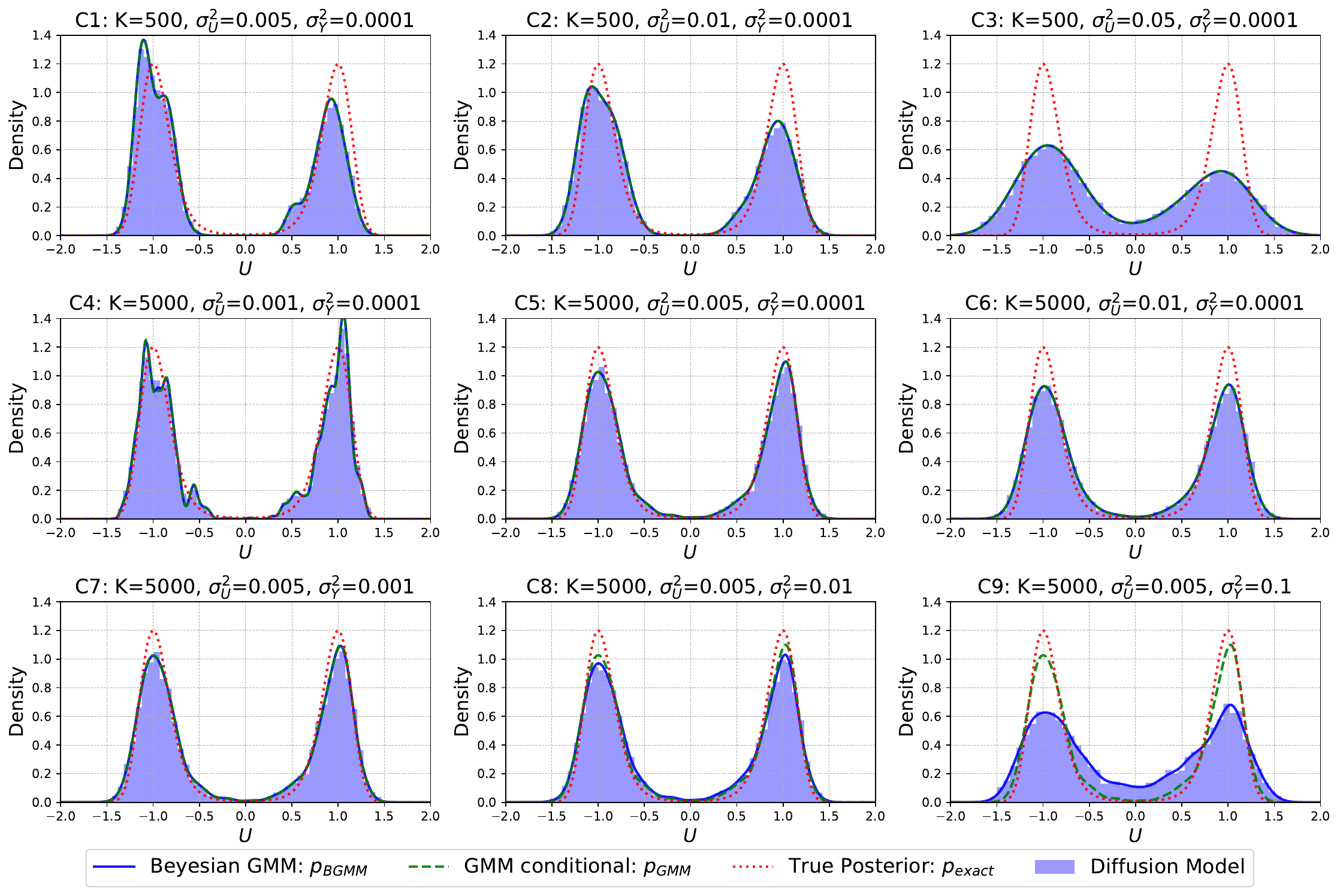}
\vspace{-0.2cm}
\caption{Density plots corresponding to each case in Table \ref{tab:full_comparison}. Across all cases, $p_{\rm BGMM}$ closely matches the diffusion model samples, indicating small discretization error and the correctness of the exact score computation. In the third row ($C_7 - C_9$), increasing $\sigma^2_{\bY}$ causes divergence from $p_{\rm GMM}$ due to Bayesian formulation error. In the first two rows ($C_1 - C_6$), increasing $\sigma^2_{\bU}$ illustrates under- and over-smoothing effects.} 
    \label{fig:rows_of_table_comparisons1-3}
\end{figure}
%
Figure~\ref{fig:rows_of_table_comparisons1-3} shows the corresponding density plots for each case in Table~\ref{tab:full_comparison}, visually confirming the trends observed in the table. Across all plots, the density $p_{\rm BGMM}$ closely matches the samples generated by the DM, indicating the small discretization error and the correctness of the exact score computation. In the third row (cases $C_7 - C_9$), we observe that increasing $\sigma^2_{\bY}$ causes the DM samples to diverge from the constructed GMM conditional distribution $p_{\rm GMM}$, highlighting the impact of the Bayesian formulation error. In the first and second rows (cases $C_1 - C_6$), $\sigma^2_{\bU}$ increases from left to right, illustrating the effects of under- and over-smoothing in approximating the true conditional distribution. Due to the small value of $\sigma^2_{\bY}$ in these cases, the diffusion model samples align well with $p_{\rm GMM}$, and the remaining discrepancy is primarily due to the mismatch between the true posterior and the GMM approximation.
\begin{figure}[h!]
    \centering
\includegraphics[width=0.5\linewidth]{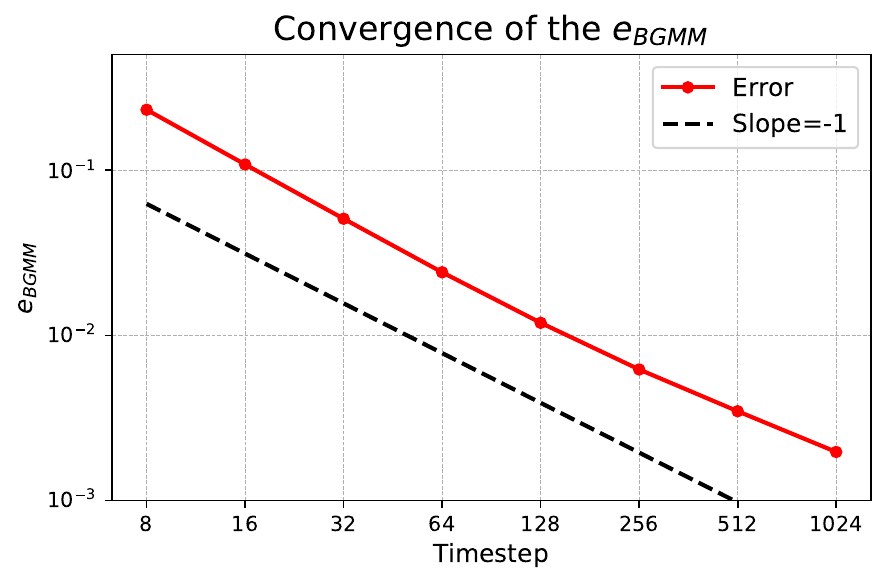}
\vspace{-0.3cm}
    \caption{
    KL divergence $e_{\rm BGMM}$ between diffusion model samples and the Bayesian posterior versus the number of time steps $N_{\tau}$ used in the reverse-time ODE. The observed linear convergence indicates the correctness of the derived conditional score.}
    \label{fig:time_error}
\end{figure}
%
To further validate the exactness of the derived conditional score expression, we examine the convergence of DM samples to the corresponding Bayesian posterior  $p_{\rm BGMM}$ as the number of time steps $N_{\tau} := 1 / \Delta \tau$  used in solving the reverse-time ODE increases. We fix 10,000 initial standard Gaussian samples and vary $N_{\tau}$ to evaluate the effect of ODE discretization. For the other parameters, we set $K = 5000$ , $\sigma^2_{\bY} = 0.0001$, and $\sigma^2_{\bU} = \sigma^2_{\bV} = 0.005$. As shown in Fig.~\ref{fig:time_error}, the KL divergence between the DM samples and $p_{\rm BGMM}$ decreases linearly with increasing $N_{\tau}$, providing further evidence for the correctness of the derived conditional score computation.

\subsection{A 20-dimensional two-mode Gaussian distribution.}\label{sec:20D_2Mode_GMM}
In this example, we apply our full proposed pipeline to the amortized conditional sampling task described in Section~\ref{sec:prob}. We consider a 20-dimensional, two-mode Gaussian mixture model (GMM) as the true joint distribution, defined as
\begin{equation}\label{eq:2mode-GMM}
f_{\bX}(\bx) = \frac{1}{2}{\mathcal{N}}(\bx; \bmu_1, \mathbf{I}_{20}) + \frac{1}{2}{\mathcal{N}}(\bx; \bmu_2, \mathbf{I}_{20})
\end{equation}
where $\bmu_1, \bmu_2 \in \mathbb{R}^{20}$ are the mean vectors of the two mixture components. Specifically, we choose $\bmu_1^\top = [1.35 \cdot \bone_5^\top, 0.5 \cdot \bone_5^\top, 0.2 \cdot \bone_5^\top, 0.1 \cdot \bone_5^\top]$, where $\bone_5 \in \bbR^5$ is a vector of ones, and set $\bmu_2 = -\bmu_1$.

To define the conditional distribution, we split the vector $\bX$ into two parts: $\bX^\top = [\bU^\top, \bV^\top]$, where $\bU \in \bbR^{d_u}$ consists of the first $d_u$ components of $\bX$ and is treated as the inference variable, and $\bV \in \bbR^{d_v}$ consists of the last $d_v$ components, viewed as the conditioning variable. The goal is to approximate the conditional distribution $p_{\bU | \bV}(\bu | \bv)$. We consider two scenarios: (i) $\bU \in \bbR^{15}$, $\bV \in \bbR^5$; and (ii) $\bU \in \bbR^{10}$, $\bV \in \bbR^{10}$. In both scenarios, we use a joint dataset $\{(\bu_k, \bv_k)\}_{k=1}^K$ of size $K = 150,000$.

To train the amortized inference network, we generate $J = 30,000$ labeled samples as defined in Eq.~\eqref{eq:labeled_dataset}, using noise parameters $\sigma^2_{\bY} = 10^{-5}$ and $\sigma^2_{\bU} = \sigma^2_{\bV} = 0.10$. The network consists of two fully connected layers with 50 neurons each and is trained using the ADAM optimizer with a learning rate of $10^{-3}$ for 50,000 epochs. Using the trained model, we generate 5,000 samples from the approximate conditional distribution and compare them with the exact conditional distribution $p_{\rm exact}$, which can be computed analytically from Eq.~\eqref{eq:2mode-GMM}.
\begin{figure}[h!]
  \centering
  \begin{subfigure}[t]{0.49\textwidth}
    \centering
    \caption{Setting (i): $\bU\in \bbR^{15},\bV\in \bbR^{5}$}
    \label{fig:20D_KDE_estimates_15D_case}
    \vspace{-0.3cm}
    \includegraphics[width=\textwidth]{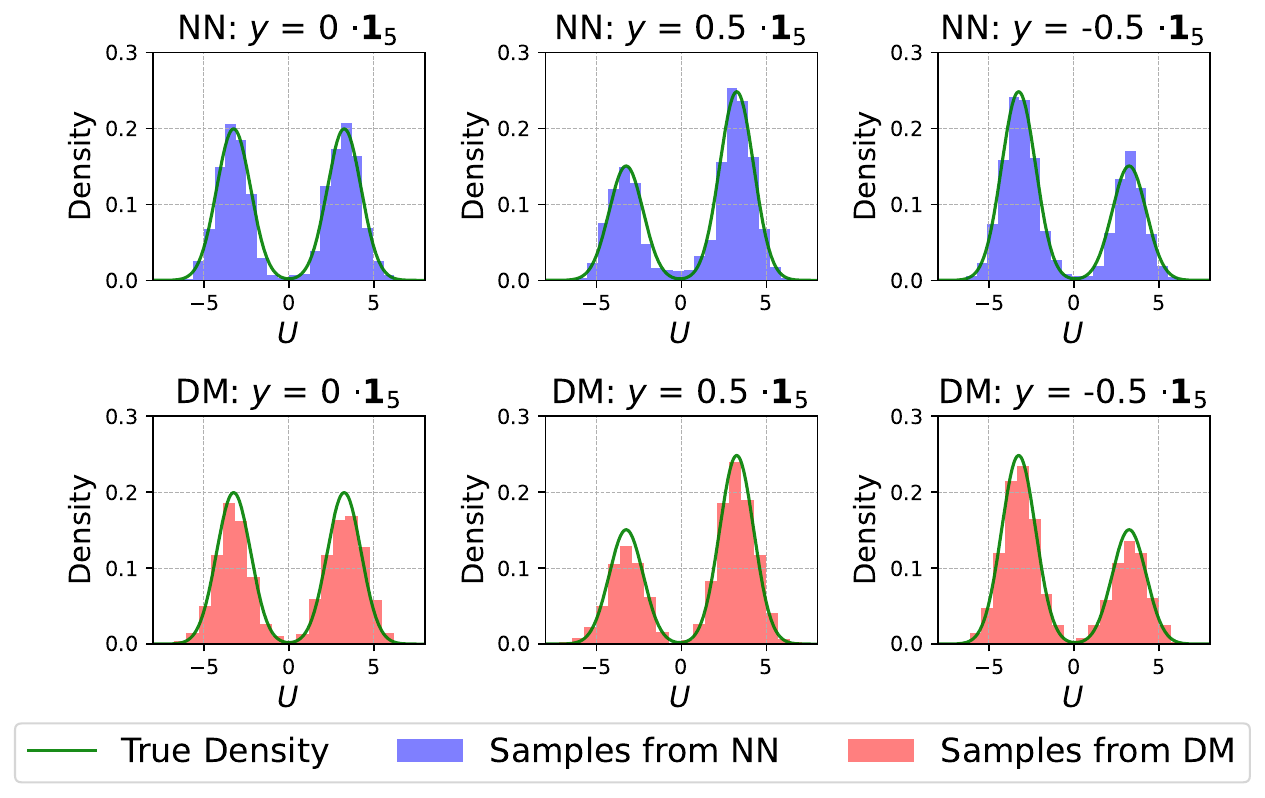} 
  \end{subfigure}
  \hfill
  \begin{subfigure}[t]{0.49\textwidth}
    \centering
    \caption{Setting (ii): $\bU\in \bbR^{10},\bV\in \bbR^{10}$}
    \label{fig:20D_KDE_estimates_10D_case}
    \vspace{-0.3cm}
    \includegraphics[width=\textwidth]{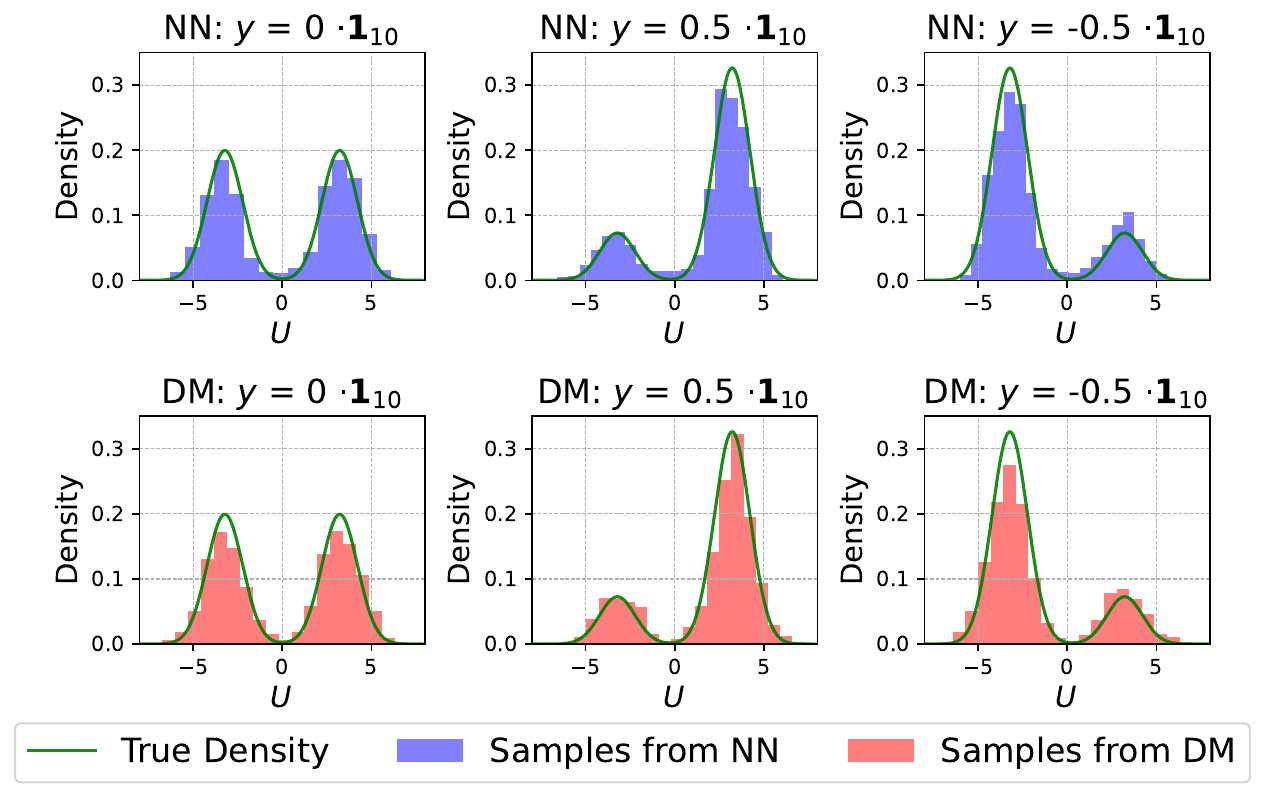} 
    \end{subfigure}
    \vspace{-0.2cm}
    \caption{
    Comparison of distributions projected onto the line connecting the mean vectors $\bmu_1$ and $\bmu_2$ for different conditioning values $\by$ under two conditional settings. In each figure, the top row shows samples generated by the diffusion model (DM), and the bottom row shows samples generated by the trained amortized inference neural network (NN). The results demonstrate strong agreement between the true conditional density and the outputs of both the DM and NN.
    }
\end{figure}
\begin{figure}[h!]
  \centering
  \begin{subfigure}[t]{0.48\textwidth}
    \centering
    \caption{Setting (i): $\bU\in \bbR^{15},\bV\in \bbR^{5}$. \\
    Conditioning value: $\by = 0 \cdot \mathbf{1}_{5}$}
    \label{fig:joint_marginals_15D}
    \vspace{-0.3cm}
    \includegraphics[width=\textwidth]{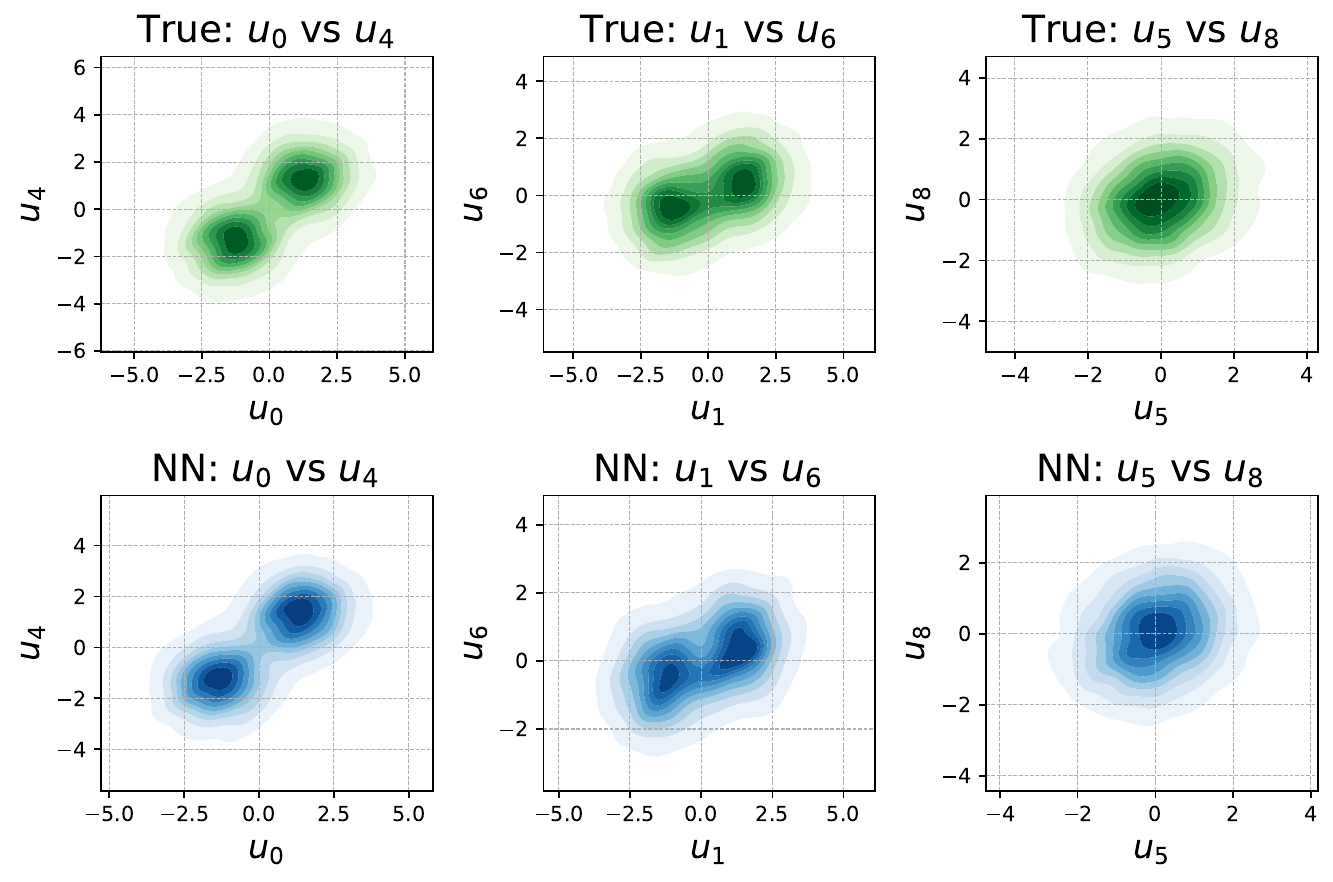} 
  \end{subfigure}
  \hfill
  \begin{subfigure}[t]{0.48\textwidth}
    \centering
    \caption{Setting (ii): $\bU\in \bbR^{10},\bV\in \bbR^{10}$.\\
    Conditioning value: $\by = 0 \cdot \mathbf{1}_{10}$.}
    \label{fig:joint_marginals_10D}
    \vspace{-0.3cm}
    \includegraphics[width=\textwidth]{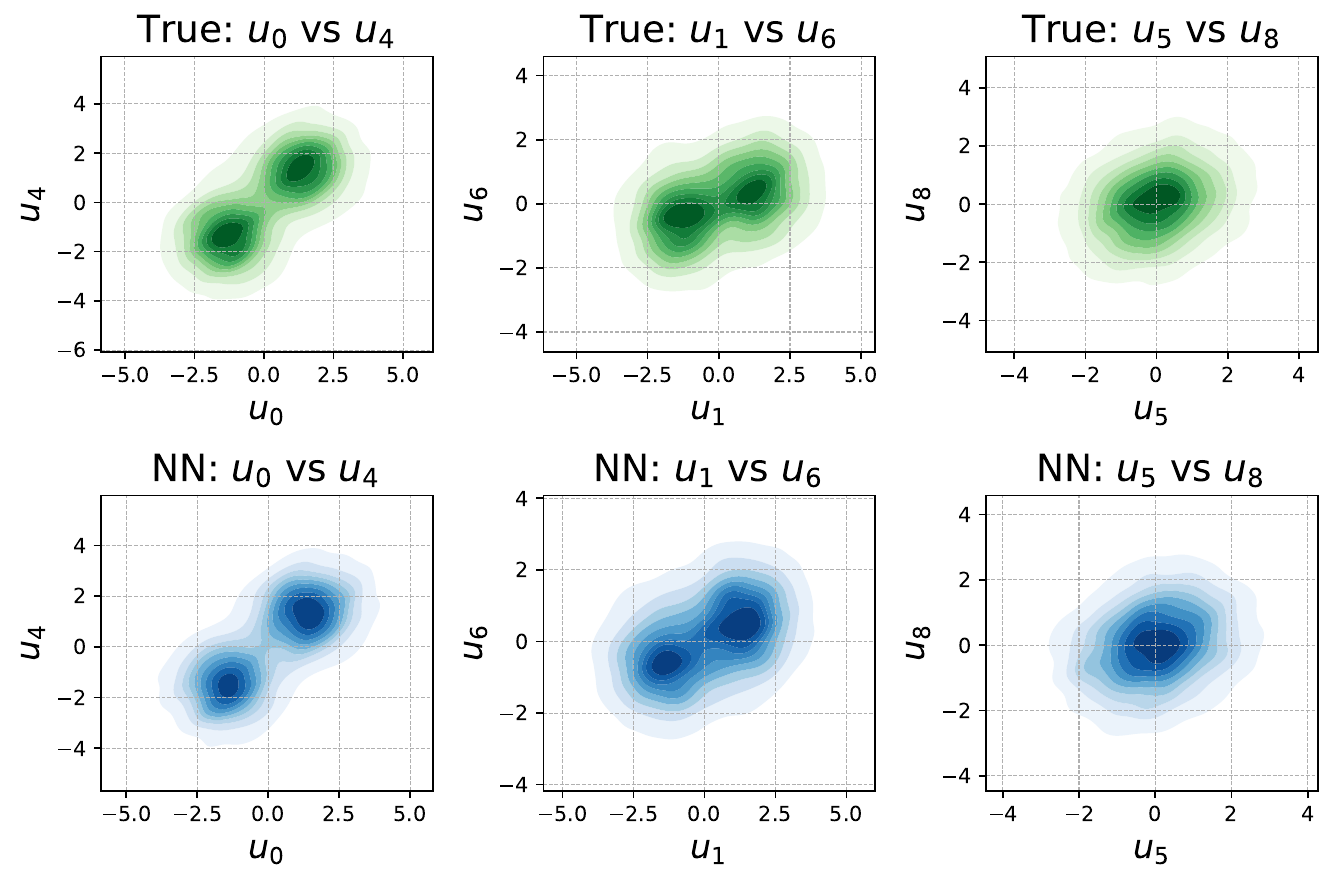} 
  \end{subfigure}
    \caption{
    Joint distribution comparison over randomly selected pairs of dimensions. The true conditional distributions are shown in green, and the corresponding distributions from the amortized inference network (NN) are shown in blue. For conditional settings (i) and (ii), the conditioning variables are $\by = 0 \cdot \mathbf{1}_{5}$ and $\by = 0 \cdot \mathbf{1}_{10}$, respectively. The results demonstrate that the NN samples accurately capture the joint structure of the conditional distributions.
    }
\end{figure}
Figures~\ref{fig:20D_KDE_estimates_15D_case} and~\ref{fig:20D_KDE_estimates_10D_case} show the conditional distributions $p_{\bU | \bV}(\bu | \bv)$ under conditional settings (i) ($\bU \in \mathbb{R}^{15}$, $\bV \in \mathbb{R}^{5}$) and (ii) ($\bU \in \mathbb{R}^{10}$, $\bV \in \mathbb{R}^{10}$), respectively, where the densities are obtained by projecting the samples onto the line connecting the two Gaussian component means. In both cases, and across varying conditioning values, we observe strong agreement between the true conditional distributions and those generated by the diffusion model (DM) and the trained amortized inference neural network (NN). This demonstrates the effectiveness of the NN in accurately approximating the conditional distribution when marginalized to a lower-dimensional space. Figures~\ref{fig:joint_marginals_15D} and~\ref{fig:joint_marginals_10D} compare the joint distributions over randomly selected pairs of dimensions. These results further confirm that the NN-based approximations not only capture marginal behavior but also preserve the joint structure of the conditional distributions.

For quantitative evaluation, Tables~\ref{tab:KL_15D1} and~\ref{tab:KL_10D1} report the KL divergence averaged over all marginal dimensions, along with the KL divergence of the projected distribution. The consistently low KL divergence values indicate strong agreement between the true conditional distributions and the generated approximations, demonstrating that the amortized inference network, which is trained using labeled data produced by the exact score-guided diffusion model, can accurately approximate complex, high-dimensional, and multi-modal conditional distributions.
\begin{table}[h!]
  \centering
  \begin{subtable}[t]{\textwidth}
    \centering
    \caption{Conditional setting (i): $\bU\in \bbR^{15},\bV\in \bbR^{5}$.
    \vspace{-0.2cm}
    }
    \label{tab:KL_15D1}
    \begin{tabular}{|c|c|c|c|}
    \hline
    \textbf{Discrepancy Type} & $\by = 0 \cdot \bone_{5}$ & $\by = -0.5 \cdot \bone_{5}$ & $\by = 0.5 \cdot \bone_{5}$ \\
    \hline
    Projection KL (NN)         & 0.0130  & 0.0156 & 0.0256 \\
    Avg. KL per dimension (NN) & 0.0040 & 0.0047 & 0.0049 \\
    \hline
    \end{tabular}
  \end{subtable}

  \vspace{0.3cm} 

  \begin{subtable}[t]{\textwidth}
    \centering
    \caption{Conditional setting (i): $\bU\in \bbR^{10},\bV\in \bbR^{10}$.
    \vspace{-0.2cm}
    }
    \label{tab:KL_10D1}
    \begin{tabular}{|c|c|c|c|}
    \hline
    \textbf{Discrepancy Type} & $\by = 0 \cdot \bone_{10}$ & $\by = -0.5\cdot \bone_{10}$ & $\by = 0.5\cdot \bone_{10}$ \\
    \hline
    Projection KL (NN) & 0.0516  & 0.0320 & 0.0304  \\
    Avg. KL per dimension (NN)           & 0.0045 & 0.0132 & 0.0160 \\
    \hline
    \end{tabular}
  \end{subtable}
  \caption{KL divergences for the 20-dimensional two-mode GMM under two conditional settings: (i) $\bU \in \mathbb{R}^{15}$, $\bV \in \mathbb{R}^{5}$, and (ii) $\bU \in \mathbb{R}^{10}$, $\bV \in \mathbb{R}^{10}$. For each case, we report the average KL divergence over all marginal dimensions and the KL divergence after projecting the distributions onto the line connecting the two Gaussian component means. Results demonstrate strong agreement between the approximate and true conditional densities in both high-dimensional settings.
  }
  \label{tab:kl_GMM}
\end{table}
\vspace{-0.7cm}

\subsection{Inference of the permeability field with uncertainty quantification.}\label{sec:elliptic_PDE}
In the final example, we consider a parameter recovery problem of the random permeability field that parameterize the following 2-dimensional elliptic PDE:
\vspace{-0.2cm}
\begin{equation}\label{eq:elliptic_pde}
\begin{aligned}
     - \nabla \cdot \left(e^{k(x,y)} \nabla u \right) &= f, \quad (x,y) \in D \\
    u(x,y) &= 0, \quad (x,y) \in \partial D
\end{aligned}
\end{equation} 
where $D = [0,1]^2$ and $f = 1$. We also assume the log-permeability field is given by the following random expansion:
\begingroup
\setlength{\abovedisplayskip}{7pt}
\setlength{\belowdisplayskip}{7pt}
\begin{equation} \label{eq:elliptic_pde_random_expansion}
    k(x,y) = \sum_{m=1}^M \sum_{l=1}^L b_{ml} \sin(2\pi mx) \sin(2\pi ly), \quad b_{ml} \sim \mathcal{N}(0,1/(m+l)),
\end{equation}
\endgroup
where we set $M=L=2$. This implies a four-dimensional coefficient vector $\bb \in \mathbb{R}^{4}$. We assume noisy observations of the solution field at $N = 10$ random spatial locations in the domain $D$, given by $\hat{u}(x) = u(x)(1 + \epsilon)$ with relative noise $\epsilon \sim \mathcal{N}(0, 0.01)$. Let $\hat{\bu} \in \mathbb{R}^{N}$ denote the vector of noisy observations. The goal is to approximate the conditional distribution $p(\bb | \hat{\bu})$, i.e., the distribution of the coefficients given the noisy and sparse observations of the solution fields.

\begin{figure}[h!]
  \centering
  \begin{subfigure}[t]{\textwidth}
    \centering
    \caption{Test case 1}
    \vspace{-0.3cm}
    \label{fig:generated_samples_PDE_testcase1}
    \includegraphics[width=0.9\textwidth]{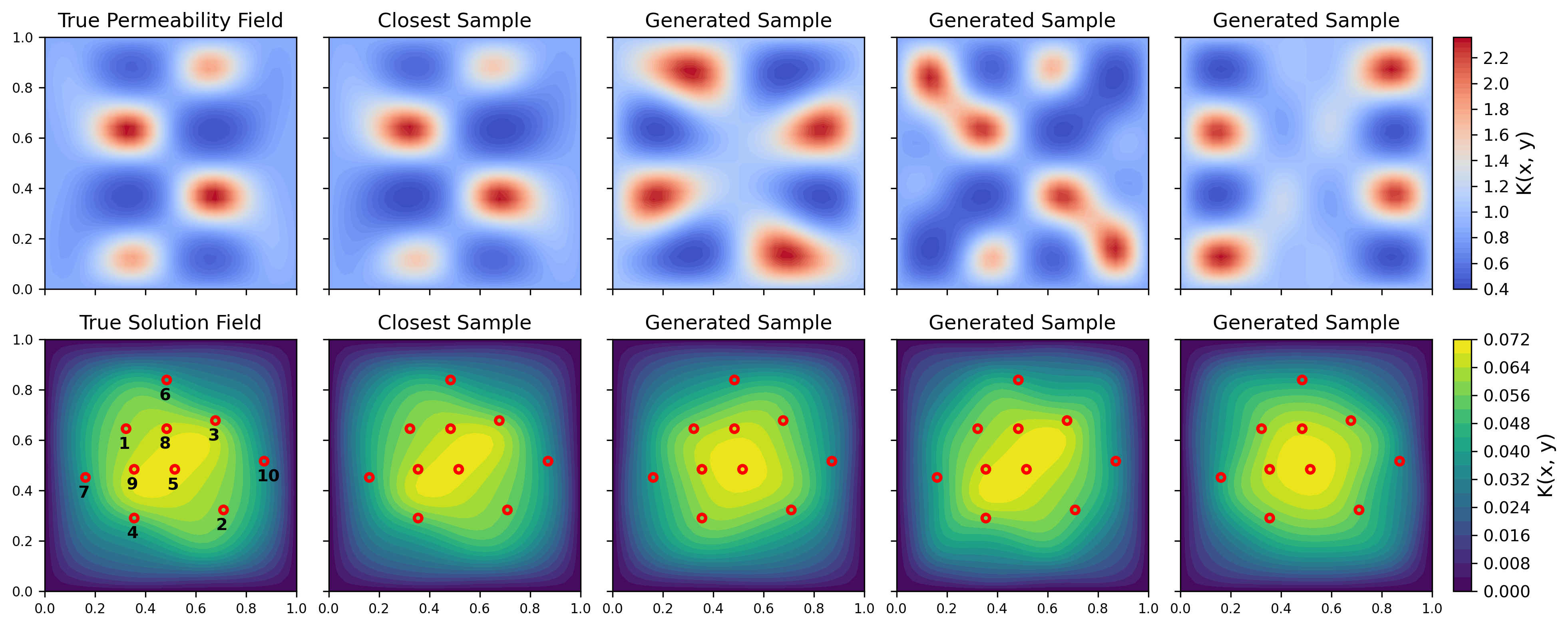} 
  \end{subfigure}
  \vspace{1em} 
  \begin{subfigure}[t]{\textwidth}
    \centering
    \caption{Test case 2}
    \vspace{-0.3cm}
    \label{fig:generated_samples_PDE_test_case2}
    \includegraphics[width=0.9\textwidth]{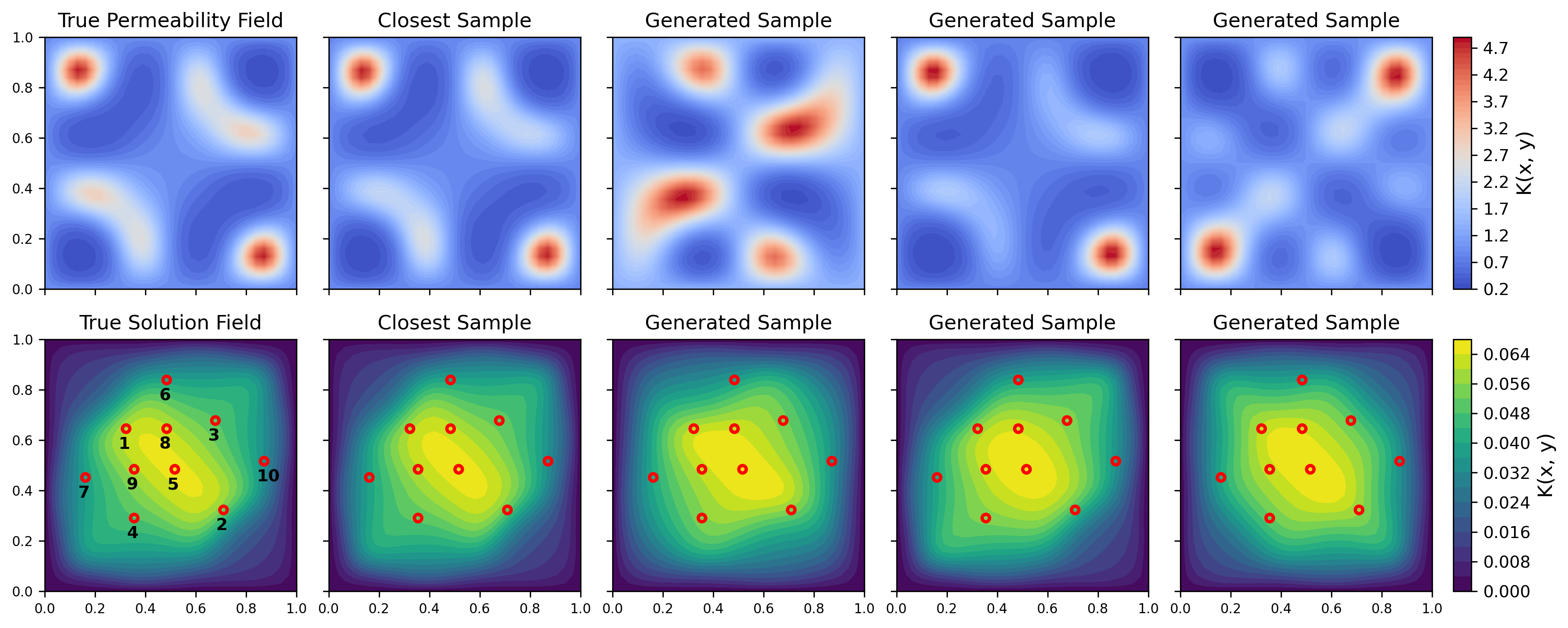} 
  \end{subfigure}
  \vspace{-0.5cm}
  \caption{
    Generated permeability and corresponding solution fields for Test Cases 1 and 2 using the amortized inference network (NN). Each case shows the ground truth (first column), the closest match (second column), and three NN-generated samples. Numbered markers indicate the 10 spatial locations used for collecting noisy solution measurements $\hat{\bu} \in \mathbb{R}^{10}$. Despite variability in the inferred permeability, the corresponding solutions closely match the ground truth, demonstrating the network’s ability to capture uncertainty while preserving predictive accuracy.
    }
    \label{fig:generated_samples_PDE_test_case_both}
\end{figure}

To obtain the joint dataset, we sample $K = 5,000$ coefficient vectors $\{\bb_k\}_{k=1}^K$ and solve Eq.~\eqref{eq:elliptic_pde} on a $32 \times 32$ grid using a second-order finite element method using those coefficients. The corresponding observed solution data $\hat{\bu}_k \in \bbR^{10}$ are then collected at 10 randomly selected spatial locations (shown in Figure~\ref{fig:generated_samples_PDE_testcase1}), forming the joint sample set $\{(\mathbf{b}_k, \hat{\bu}_k)\}_{k=1}^{K}$. 
We use this joint dataset to construct the prior GMM, with variance parameters set to $\sigma^2_{\bU} = \sigma^2_{\bV} = 0.10$ and $\sigma^2_{\bY} = 10^{-5}$. The reverse-time ODE is solved using $1000$ time steps. To train the amortized inference network, we use the exact conditional-score diffusion model to generate a labeled dataset of size $J = 5000$, as defined in Eq.~\eqref{eq:labeled_dataset}. The network consists of a single hidden layer with 100 neurons and is trained using the ADAM optimizer with a learning rate of $10^{-3}$ for $20,000$ epochs.
After training, we evaluate the performance of the amortized inference network by generating conditional distributions $p(\bb | \hat{\bu})$ based on two different noisy solution inputs $\hat{\bu}$, referred to as Test Case 1 and Test Case 2 in all reported figures. For each test case, we generate $5000$ samples from the approximate conditional distribution to assess the quality of the inference.
%

Figures~\ref{fig:generated_samples_PDE_testcase1} and~\ref{fig:generated_samples_PDE_test_case2} show the permeability field samples generated by the amortized inference network, along with the corresponding solution field samples. Permeability fields are reconstructed from the expansion in Eq.~\eqref{eq:elliptic_pde_random_expansion} with $\bb$ produced by NN, and the solution fields are obtained by solving the PDE with the resulting permeability fields. The results demonstrate that, although the generated permeability fields exhibit significant variability, the resulting solution fields match the true solution closely. This highlights the model’s ability to capture uncertainty in the inferred permeability while maintaining accuracy in the predicted solution. For comparison, the standard nearest neighbor method may produce point estimates that fit the observed data well but fail to capture the underlying uncertainty in the parameter space.
\begin{figure}[h!]
    \centering
\includegraphics[width=\linewidth]{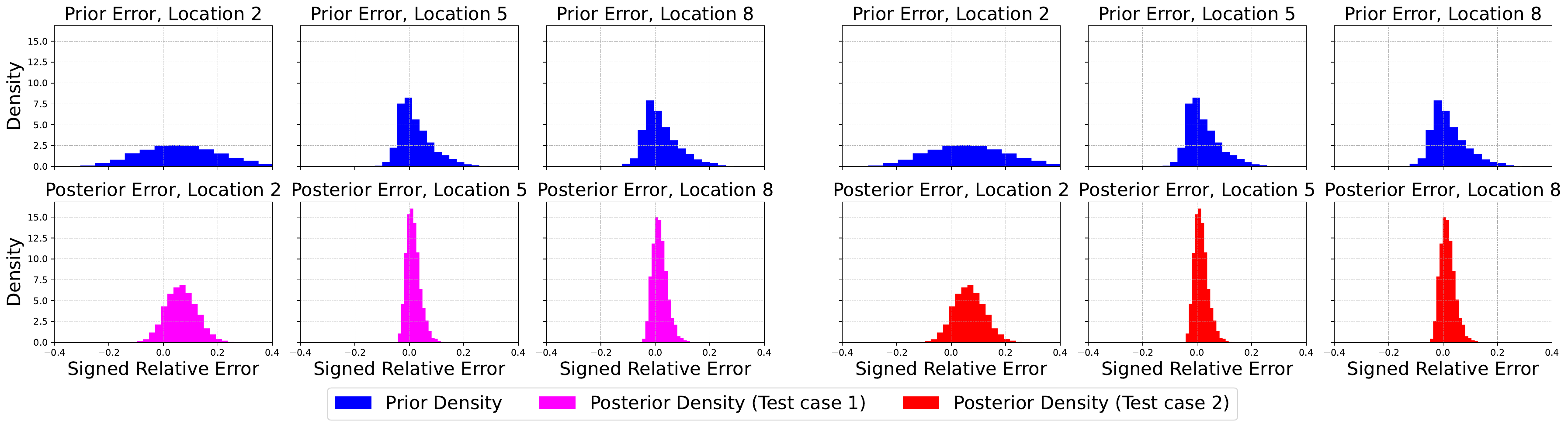}
\vspace{-0.5cm}
    \caption{Signed relative error distributions at three sample locations, before (top) and after (bottom) conditional sampling from the amortized inference network. The reduced spread indicates improved alignment with the observed solutions, and the remaining variability reflects inherent uncertainty from measurement noise.
    }
    \label{fig:signed_relative_error2}
\end{figure}
%

To further quantify the uncertainty for the amortized inference network, we examine the relative error distribution of the solution field at three sample locations from Figure~\ref{fig:generated_samples_PDE_test_case_both}, both before and after conditioning. As shown in Figure~\ref{fig:signed_relative_error2}, the spread of the relative error is significantly reduced after conditioning, indicating that the amortized inference network effectively infers the expansion coefficients $\bb$ in a way that incorporates the observational data and aligns with the true solution field. The remaining variability in the posterior distribution reflects inherent uncertainty from measurement noise, which is also preserved by the network.
\begin{figure}[h!]
    \centering
    \includegraphics[width=\linewidth]{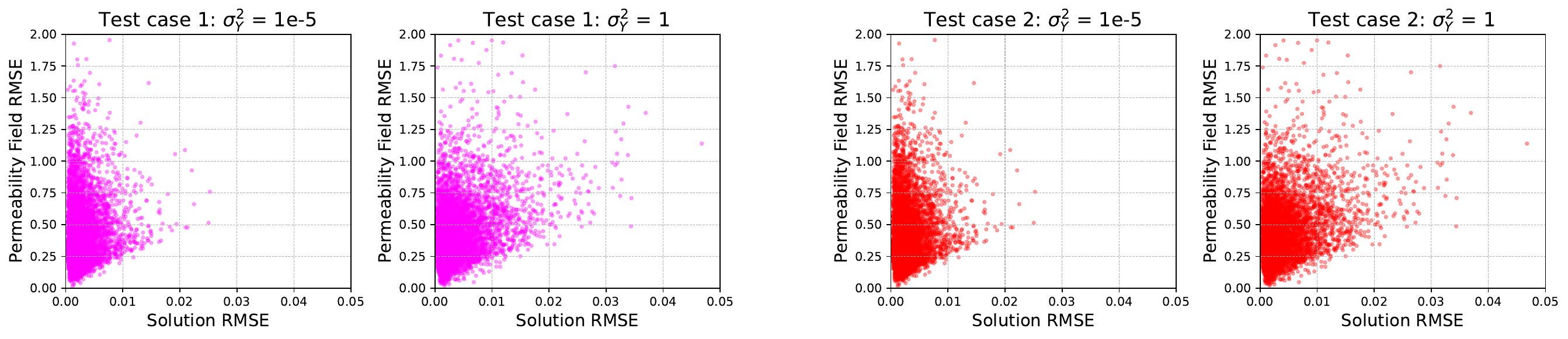}
    \vspace{-0.5cm}
    \caption{
    Joint distribution of relative MSE in generated permeability and solution fields for two test cases and two $\sigma^2_{\bY}$ values. Permeability fields show greater variability than solution fields, indicating that the model preserves parameter uncertainty while aligning with observed data. Results also highlight the ill-posed nature of the inference problem and the over-smoothing effect from a large value of $\sigma^2_{\bY}$ to the posterior.
    }
    \label{fig:MSE_relative_error}
\end{figure}
%

In Figure~\ref{fig:MSE_relative_error}, we show the joint distribution of the relative MSE of the generated permeability fields versus the corresponding solution fields for two test cases and two different $\sigma^2_{\bY}$ values. 
In both test cases, the permeability fields exhibit significantly greater variability than the solution fields, indicating that the generated conditional samples align well with the observed solution data while still preserving meaningful uncertainty in the permeability fields. 
The concentration of errors in the upper half of the plots highlights the ill-posedness of the inference problem, where small changes in the solution field correspond to large changes in the permeability, but a relatively stable forward problem. This asymmetry underscores the importance of performing uncertainty quantification when solving such inference problems.
Additionally, larger values of $\sigma^2_{\bY}$ introduce greater uncertainty due to over-smoothing, leading to increased spread in both the solution error and the inferred permeability. This effect can be mitigated by choosing a smaller $\sigma^2_{\bY}$ to better preserve fidelity to the observed data.

\begin{figure}[htbp]
    \centering
\includegraphics[width=0.49\linewidth]{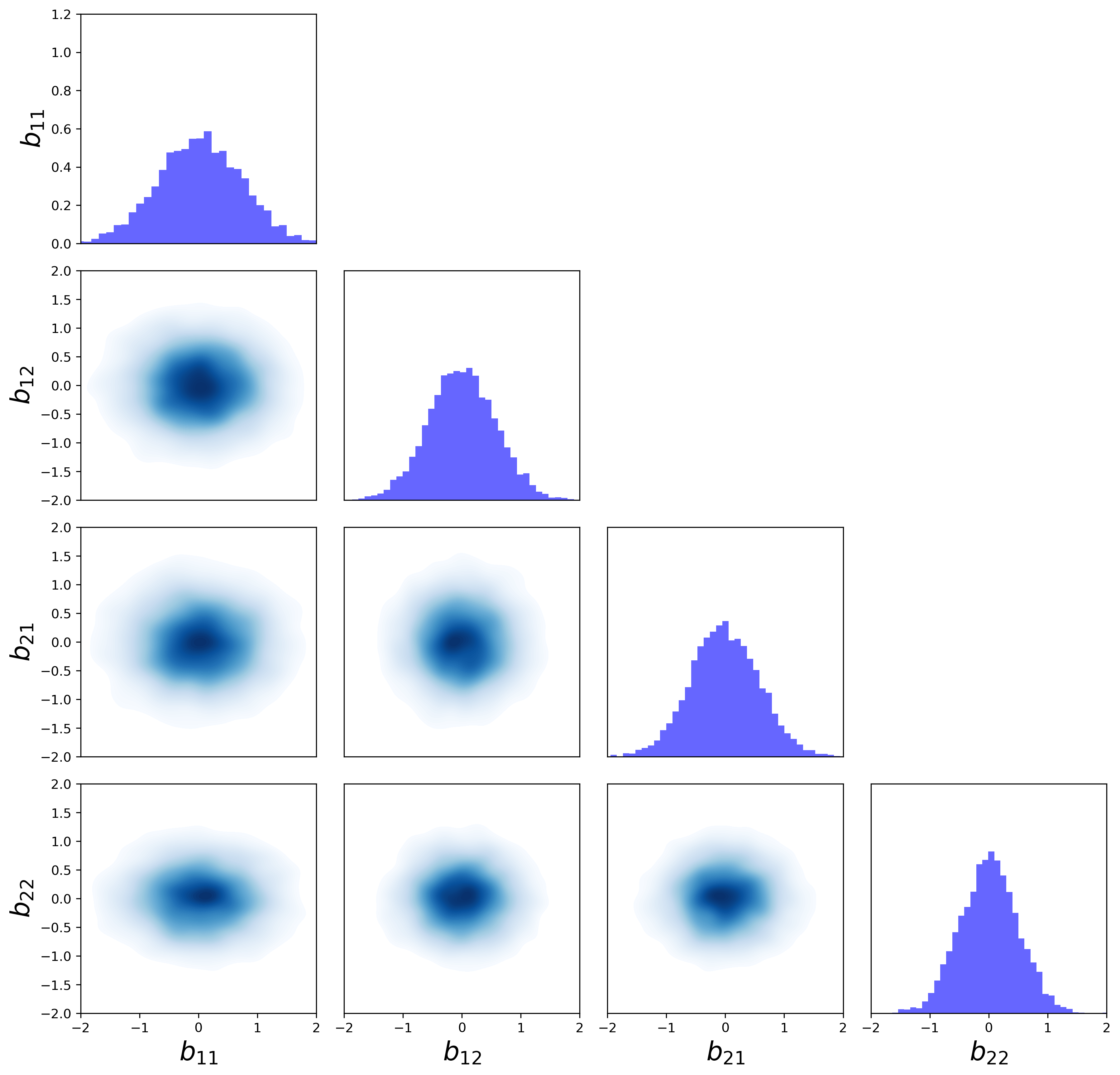}
\includegraphics[width=0.49\linewidth]{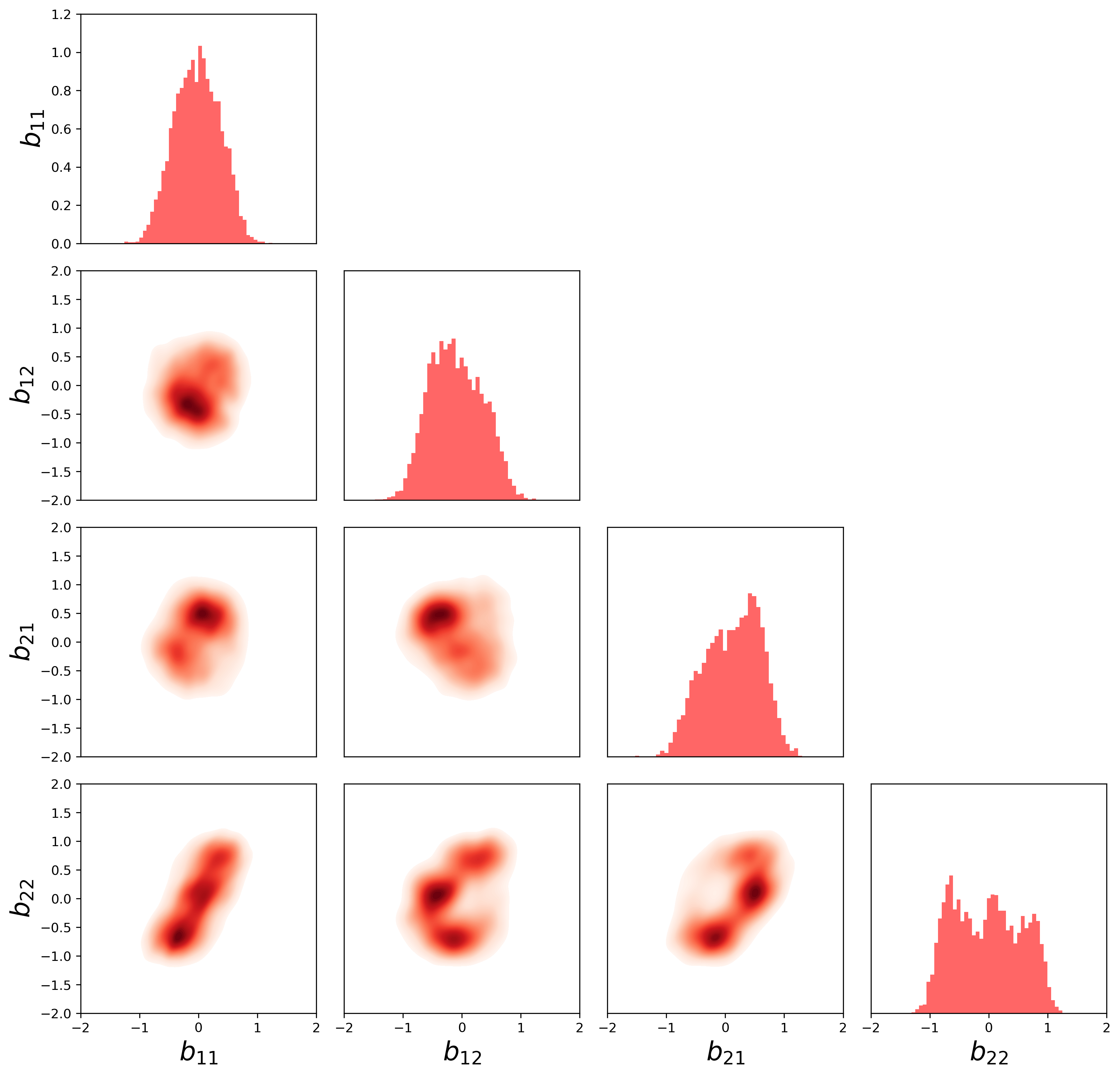}
    \caption{
    Pairwise joint distributions of the inferred coefficients $\bb = [b_{11}, b_{12}, b_{21}, b_{22}]^\top$ for Test Case 2. The left panel (blue) shows the prior, and the right panel (red) shows the posterior. The posteriors show clear deviations from the prior, reflecting the impact of conditional sampling to match the observed solution data.
    }
    \label{fig:coefficient_posterior}
\end{figure}
%

Figure~\ref{fig:coefficient_posterior} shows the pairwise joint distributions of the inferred coefficients $\bb$ for Test Case 2. Comparing the prior and posterior distributions reveals noticeable changes in the posterior, reflecting the impact of conditioning on the observed solution data. The proposed method successfully generates a diverse set of coefficient samples that, when used to solve the underlying PDE, yield solutions with low error at the observed spatial locations.

\section{Conclusion} In this work, we have presented an exact conditional score-guided diffusion model for generating samples from the target conditional distributions. In addition, we proposed a supervised training framework of an amortized Bayesian inference network that enables efficient posterior sampling conditioned on various input observation values. Approximating high-dimensional and multi-modal conditional distributions remains a central challenge in the scientific community, particularly when only sparse prior data are available. By analytically deriving the conditional score function under a Gaussian Mixture Model (GMM) prior, we can obtain a tractable approximation to the target conditional distribution without requiring the score network training. In our numerical studies, we assume spherical covariance structures and demonstrate successful uncertainty quantification in multi-modal synthetic problems and a parameter estimation task. As a direction for future work, we plan to extend the framework by exploring anisotropic covariance structures to enable more expressive GMM prior and generalizing the exact conditional score to handle nonlinear observation operators $h(\bx)$, beyond the current linear case $\bH \bx$. Overall, our analytically derived, training-free conditional score function combined with the amortized Bayesian inference network offers a principled and efficient approach to score-based Bayesian inference and amortized conditional sampling.

\section*{Acknowledgment.}
The work of Z.~Zhang and G.~Zhang was partially supported by the U.S. Department of Energy, Office of Science, Office of Advanced Scientific Computing Research, Applied Mathematics program, under the contracts ERKJ388 and ERKJ443. 
ORNL is operated by UT-Battelle, LLC., for the U.S. Department of Energy under Contract DE-AC05-00OR22725. The work of C.~Tatsuoka and D.~Xiu was partially supported by AFOSR FA9550-22-1-0011.

\bibliographystyle{siamplain}
\bibliography{references,GZ_ref}

\end{document}